\documentclass[twocolumn,showpacs,preprintnumbers,amsmath,amssymb,aps,pra,floatfix]{revtex4}

\usepackage{graphicx}
\usepackage{epsf}
\usepackage{pstricks}
\usepackage{graphicx}
\usepackage{dcolumn}
\usepackage{bm}

\begin{document}

\title{Low-energy excitations of a boson pair in a double-well trap}

\author{D. S.  Murphy}
\author{J. F. McCann}
\email{j.f.mccann@qub.ac.uk} \affiliation{
Centre for Theoretical Atomic, Molecular and Optical Physics, The Queen's 
University Belfast, Belfast, BT7 1NN, UK}%

\date{\today}

\begin{abstract}
The states of a boson pair in a one-dimensional double-well potential are 
investigated.  Properties of the ground and lowest excited states of 
this system are studied, including the two-particle wavefunction, momentum pair
distribution and entanglement.  The effects of varying both the barrier height
and the effective interaction strength are investigated.
\end{abstract}

%
%

\maketitle

\section{\label{sect:intro} Introduction}
Ensembles of ultracold, trapped atoms provide an ideal test system for the study
of fundamental quantum principles.  The manipulation of atoms with photons,
\cite{coh98}, has given rise to the experimental realization of Bose-Einstein
condensation (BEC) \cite{and95, dav95, dal99} and, more recently, the trapping
and manipulation of condensates using optical lattice potentials 
\cite{and98, cat01, bloA05, bloB05}.  The weak coupling of neutral atoms to 
their environment mean that this system of cold neutral atoms, confined by a 
periodic potential, may prove useful in the investigation of primitive quantum 
information processing \cite{mon02}.  Indeed, such systems have already been 
used to carry out a two-qubit entangling operation \cite{jak99, man03}, thereby 
realizing the crucial CNOT gate.  At the same time, the spatially periodic 
nature of the system makes it ideal for the detailed study of solid-state 
Hamiltonians \cite{fis89, jak98, gre02}.  The benefit of this artificial system,
in this regard, lies in the fact that the experimentalist can easily vary 
external control parameters (e.g. laser intensity or wavelength), thereby 
varying particular parameters of the system Hamiltonian.  A degree of control 
that is not generally afforded to typical solid-state systems.

The dynamics of a system of ultracold atoms, confined by an optical lattice
potential, can be accurately described within the framework of the Bose-Hubbard
model \cite{fis89,jak98}.  In this model the system Hamiltonian is parameterized
by the tunnelling strength between adjacent lattice sites, $J$, and the on-site
interaction energy, $U$.  The Hamiltonian describing the system dynamics can
then be written as 
 \begin{equation}
 \label{eq:bose_hubbard_ham}
    \hat{H} = J \sum_{\left< i,j \right>} \hat{b}_{i}^{\dagger} \hat{b}_{j} \;
    + \; \sum_{i} \epsilon_{i} \hat{n}_{i} \; + \; U \sum_{i}
    \hat{n}_{i} \left( \hat{n}_{i} - 1 \right) \hspace*{0.3cm} ,
 \end{equation} 
where $\hat{b}_{i}^{\left( \dagger \right)}$ is the annihilation (creation)
operator for an atom at the lattice site $i$ and $\hat{n}_{i} =
\hat{b}_{i}^{\dagger} \hat{b}_{i}$ is the number operator for that site. 
Parameter $\epsilon_{i}$ is the single-particle energy at lattice site $i$ and
will vary with $i$ for an inhomogeneous lattice.  Implicit in this model is the 
assumption that the dynamics of the system is dominated by single- and 
two-particle effects.  In this way, the system of two, confined, interacting 
particles represents the fundamental building block for the understanding of 
these many-body systems.  Furthermore, continual advancement in optical lattice 
technology means that it has become possible to confine small numbers of atoms 
(e.g. 1 or 2) at individual lattice sites, effectively realizing a system of 
two trapped atoms.

For low-energy collisions the particle interactions can be accurately 
represented within the pseudopotential approximation \cite{hua_st}.  The 
eigenstates for a system of two particles, interacting via a pseudopotential, 
can be determined analytically for both isotropic \cite{bus98} and anisotropic 
\cite{idz05} harmonic traps.  Under such confinement, the `free-space' 
pseudopotential approximation is found to be sufficiently accurate provided the
length scale associated with the particle-particle interactions ($a$) is short 
compared to the length scale of the confining potential ($L$) \cite{blo02}.  
For the case in which $a$ and $L$ are comparable, one may introduce an 
energy-dependent scattering length and solve for the eigenenergies of the 
system self-consistently \cite{tie00, bol02, bol03}.

In addition to providing small numbers of particles at individual lattice sites,
optical lattice experiments also allow for the realization of quasi-one and -two
dimensional systems \cite{kin04, par04}.  Simply increasing the confining 
potential steeply in one or two of the transverse directions will effectively 
`freeze out' the corresponding degrees of freedom \cite{ket96, gor01}.  Such 
systems of reduced dimensionality can also be achieved using optical or magnetic
atom waveguides.  The theoretical treatment of the particle-particle
interactions in such low-dimensional geometries has been previously considered. 
For a quasi-one dimensional (quasi-1D) system it was found that the scattering 
could be treated in terms of a 1D, zero-ranged $\delta$-potential, renormalized 
according to the confining potential \cite{ols98}.  The physical realization of 
such quasi-1D trap geometries and recent advances in the tuning of atomic 
interactions using Feshbach resonances have permitted the study of previously 
inaccessible regimes.  Notably, the 1D system of impenetrable bosons, or so-called 
Tonks-Girardeau gas \cite{ton36, gir60}, has commanded considerable experimental 
\cite{kin04, par04} and theoretical \cite{yuk05, gir00, gir01, bus03, lin07, 
murA07} interest in recent years.

In \cite{murA07} the detailed theoretical study of two interacting particles in
a $\delta$-split harmonic potential was considered.
The DVR techniques, \cite{bay86, lig00}, employed in \cite{murA07} to study the 
$\delta$-split trap potential can be easily adapted to other types of confining 
potentials.  In the current article we utilize these same numerical techniques 
to study a prototypical two-well trap, defined by $  V\left( x \right) = A 
\left[ x^{4} - \kappa x^{2} \right]$.  The eigenspectrum for this two-particle 
system is studied and properties of the ground and lowest excited states are 
investigated for varying of the barrier height (dictated by $\kappa$) and the 
strength of the particle-particle interactions.  Particular consideration is 
given to the similarities observed between the the ground state structure in 
this prototypical two-well potential and that of the $\delta$-split potential 
\cite{murA07}.

Similar numerical studies of ultracold few-boson systems have been recently
reported \cite{zolA06, zolB06, zol07}.  In this work the authors use a
multi-configurational time-dependent Hartree (MCTDH) method to study systems of 
several bosons in a double-well trap, with narrow width Gaussians used to 
represent both the central splitting potential and the interparticle potential.  
Where comparison is possible, the results of this numerical MCTDH study 
demonstrate qualitative similarity to the results of the present study. 

The remainder of this paper is organized as follows.  In Sec. 
\ref{sect:system_hamiltonian} we outline the Hamiltonian that shall be 
considered, for two particles confined by a quasi-1D double well potential.  
In Sec. \ref{sect:energy_spec} we present the energy level spectrum for the
single and two-particle systems, illustrating how the spectrum is influenced by
barrier height and interaction strength.  In Sec. \ref{sect:ground_state} we 
examine various properties of the two-particle ground state.  The properties
considered include the ground state wavefunction, momentum distributions (Sec.
\ref{subsect:mom_dist_ground}) and von Neumann entropy (Sec.
\ref{subsect:entropy_ground}).  Particular emphasis is given to how these
properties may be influenced by varying the `experimentally controllable' 
parameters of barrier height and interaction strength.  In Sec.
\ref{sect:excited_states} we systematically examine these same properties for the
lowest excited states of this system.  Finally, in Sec. \ref{sect:summary} we
summarize our findings and make some concluding remarks.

\section{\label{sect:system_hamiltonian} System Hamiltonian}
Consider a system of two interacting particles confined in two dimensions by 
means of a `tight' harmonic potential, having trapping frequency
$\omega_{\perp}$ and associated length scale $d_{\perp} = 
\sqrt{\hbar/m\omega_{\perp}}$.  In the remaining third dimension, the confining 
potential is, relatively, `loose' and has the form
   \begin{equation}
   \label{eq:double_well_unscaled}
      V \left( x \right) = A \left[ x^{4} - \kappa x^{2} \right]  
      \hspace*{0.3cm} .
   \end{equation}
The parameters $A$ and $\kappa$ determine the precise form of the double-well
potential.  It is straightforward to verify that the two minima of this
double-well potential are located at $ x_{\textrm{min}} =  \pm \sqrt{\kappa/2}$,
with the potential at these minima being $ V \left( x_{\textrm{min}} \right) = 
-A \kappa^{2}/4$.  The well separation, $x_{\textrm{min}}$, and the barrier 
height, $V \left( x_{\textrm{min}} \right)$, are controlled by the parameter 
$\kappa$.   

As a result of the large energy level separation, associated with the 
transverse eigenstates ($\hbar \omega_{\perp}$), the transverse motion of the
particles is `frozen out'.  In this way the particles are confined to the lowest
motional state in each of these transverse directions.  In this case the system 
is quasi-1D and may be effectively described by
   \begin{equation}
   \label{eq:two_part_ham_unscaled}
       H = \sum_{i = 1,2} \left[- \frac{\hbar^{2}}{2m}
       \frac{\partial^{2}}{\partial x_{i}^{2}} + A \left( x_{i}^{4} - \kappa
       x_{i}^{2} \right) \right] + g_{1D} \delta \left( x_{2} - x_{1} \right)
       \hspace*{0.3cm} . 
   \end{equation}
Here, $m$ is the mass, and $x_{1}$ and $x_{2}$ are the coordinates of atoms 1 
and 2, respectively.  The quantity $g_{1D}$ represents the particle-particle 
interaction strength, and is related to the 1D s-wave scattering length 
($a_{1D}$) through $ \hspace*{0.1cm} g_{1D} = -2\hbar^{2}/m a_{1D}
\hspace*{0.1cm}$.  In turn, $a_{1D}$ is related to the 3D s-wave scattering 
length, $a_{3D}$, through $\hspace*{0.1cm} a_{1D} = -d_{\perp}^{2}/2a_{3D}(1 -
Ca_{3D}/d_{\perp}) \hspace*{0.1cm}$, where $C$ is a constant and has 
approximate value $C = 1.4603$ \cite{ols98}.   

In the limit of tight confinement the free-space pseudopotential approximation,
for the particle-particle interactions, becomes compromised \cite{blo02, tie00}.
In this case, one may obtain the eigenenergies for the system by employing an 
energy-dependent scattering length and solving for the energy eigenvalues 
self-consistently \cite{bol02, bol03, bur02}.  For current purposes it is 
supposed that we are in the regime for which the pseudopotential approximation 
is still valid and the 1D collisional coupling, $g_{1D}$, acts as a parameter 
for the system.  

The aim is to study the eigenvalues and eigenvectors for the 2D Hamiltonian
given in Eq. (\ref{eq:two_part_ham_unscaled}). To facilitate this we introduce 
the scaling $x_{i} = \alpha \bar{x_{i}}$ for $i = 1, 2$.  Under this rescaling 
the time-independent Schr\"odinger equation (TISE) can be written as 
   \begin{equation} 
   \label{eq:tise_scaled}
      \bar{H} \Psi_{i} \left( \bar{x_{1}}, \bar{x_{2}} \right) = \bar{E_{i}} 
      \Psi_{i} \left( \bar{x_{1}}, \bar{x_{2}} \right) \hspace*{0.3cm} ,
   \end{equation}
where
   \begin{equation} 
   \label{eq:ham_scaled}   
      \bar{H} = \frac{m \alpha^{2}}{\hbar^{2}} H = \sum_{i = 1,2} \left[ 
      \frac{1}{2} \frac{\partial^{2}}{\partial \bar{x}_{i}^{2}} + \left( 
      \bar{x}_{i}^{4} - \bar{\kappa} \bar{x}_{i}^{2} \right) \right] + 
      \bar{g}_{1D} \delta \left( \bar{x}_{2} - \bar{x}_{1} \right) 
      \hspace*{0.3cm} . 
   \end{equation}
Here the scaling factor, $\alpha$, has been chosen such that 
   $$
      \frac{A m \alpha^{6}}{\hbar^{2}} = 1 \hspace*{0.3cm} .
   $$
Consequently,
   \begin{equation}
   \label{eq:scaled_quantities}
      \begin{array}{rcl}
         \bar{\kappa} & = & \left( \frac{A m}{\hbar^{2}} \right)^{1/3} \kappa
         \hspace*{0.3cm} , \\[0.2cm]
	 \bar{g}_{1D} & = & \frac{m}{\hbar^{2}} \left( \frac{\hbar^{2}}{A m} 
	 \right)^{1/6} g_{1D} \hspace*{0.3cm} \textrm{and} \\[0.2cm]
	 \bar{E} & = & \frac{m}{\hbar^{2}} \left( \frac{\hbar^{2}}{A m}
	 \right)^{1/3} E \hspace*{0.3cm} .
      \end{array} 
   \end{equation}
For convenience we shall drop `bar' on all quantities and use, exclusively, the
scaled quantities just described.

\section{\label{sect:energy_spec} Energy spectrum}
The eigenspectra for the single- and two-particle system are obtained, subject 
to the scaling introduced in the previous section.  A cartesian DVR \cite{bay86}
is used to discretize the spatial coordinates $x_{1}$ and $x_{2}$, see 
\cite{murA07} for details.  The discretization scheme used in these calculations
employs $N = 61$ mesh points in each dimension with a mesh spacing of $h = 0.16$.
Consideration is mainly limited to the four lowest eigenvalues of the
two-particle system (i.e. the lowest band), in particular we are interested in
the behavior as the parameters $\kappa$ and $g_{1D}$ are varied.

\subsection{\label{subsect:sing_part_spec} Single-particle spectrum} 
\begin{figure}[!ht]
      \begin{center}  
         \includegraphics[scale = 0.3]{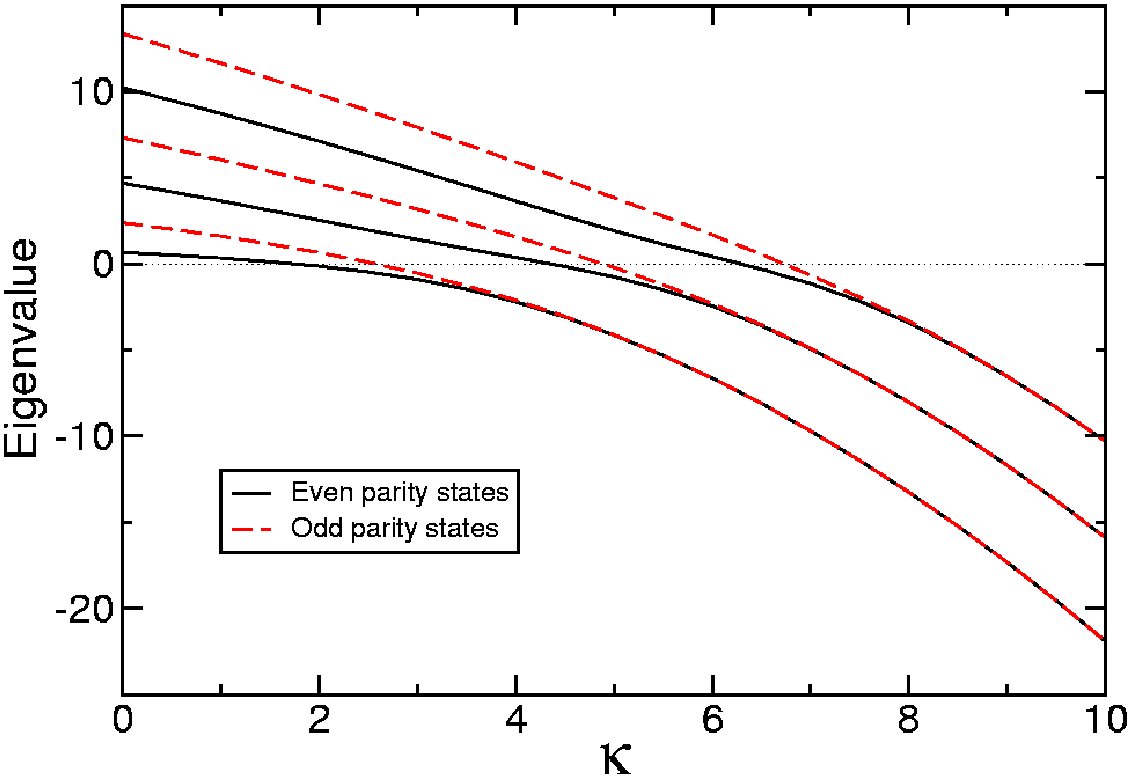} 
      \caption{\label{fig:sin_part_eig_spec} Lowest energy eigenvalues for a
      single particle in the potential $V \left( x \right) = x^{4} - \kappa
      x^{2}$.  The eigenstates alternate between states of odd parity 
      (dashed lines) and even parity (solid lines).  As the barrier is
      introduced these states pair up.}
      \end{center}
   \end{figure}

Subject to the scaling introduced in Sec. \ref{sect:system_hamiltonian} the
TISE for the single-particle system is simply
   \begin{equation}
   \label{eq:sing_part_ham_scaled}
      \left[ - \frac{1}{2} \frac{\partial^{2}}{\partial x^{2}} + \left( x^{4} -
      \kappa x^{2} \right) \right] u_{i} \left( \kappa ; x \right)  =
      E^{\textrm{single}}_{i} \left( \kappa \right)  u_{i} \left( \kappa ; x 
      \right) \hspace*{0.3cm} .
   \end{equation}
The single-particle spectrum is presented in Fig. \ref{fig:sin_part_eig_spec}. 
One can see that as the parameter $\kappa$ is increased the lowest eigenvalues
are pulled downwards in energy as $V \left( x_{\textrm{min}} \right)$ becomes
increasingly negative.  At the same time one observes the degeneracy of energy 
levels as $\kappa$ is increased.  

\subsection{\label{subsect:two_part_spec} Two-particle spectrum} 
   \begin{figure*}[!ht]
      \begin{center}  
         \includegraphics[scale = 0.5]{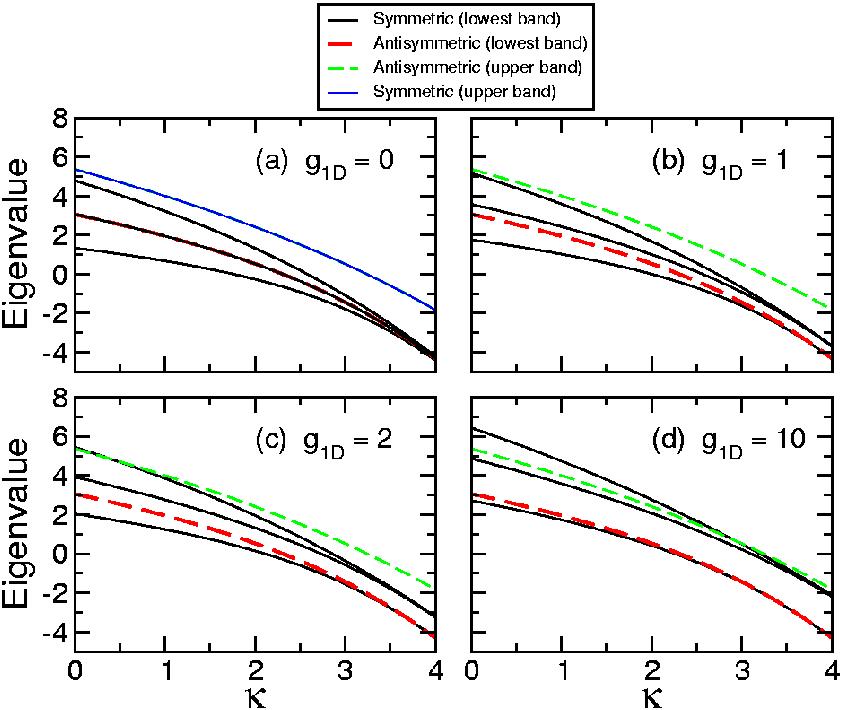} 
      \caption{\label{fig:two_part_eig_spec} Lowest energy eigenvalues for the
      system of particles in a double-well potential of the form $V \left( x 
      \right) = x^{4} - \kappa x^{2}$, as a function of the barrier height,
       $\kappa$.  The spectrum is displayed for four different values of
       interaction strength: $g_{1D} = 0$ (a), $g_{1D} = 1$ (b), $g_{1D} = 2$
       (c) and $g_{1D} = 10$ (d).  Introduction of the interaction coupling has 
       the effect of shifting the symmetric states (solid lines) upwards in 
       energy, whilst the antisymmetric states (dashed lines) remain 
       unaffected.}
      \end{center}
   \end{figure*}
Extending consideration to the two-particle spectrum, we focus attention on the
two-particle eigenstates belonging to the lowest band.  Denoting the
$i^{\textrm{th}}$ eigenstate of the two-particle system by $\Psi_{i}$, the 
eigenstates for the lowest band are then denoted by $\Psi_{0}, \Psi_{1},
\Psi_{2}$ and $\Psi_{3}$ (see Sec. \ref{sect:excited_states} for details).  This
lowest band corresponds to the four lowest levels in Fig. 
\ref{fig:two_part_eig_spec}(a), representing the two-particle system in the 
absence of interactions.  In this non-interacting regime the two-particle 
eigenstates, under exchange symmetry, are
   \begin{align}
      \Psi^{\textrm{ni}}_{0} \left( \kappa ; x_{1}, x_{2} \right) = & u_{0} 
      \left( \kappa ; x_{1} \right) u_{0} \left( \kappa ; x_{2} \right) 
      \hspace*{0.3cm} , \nonumber \\[0.2cm]
      \Psi^{\textrm{ni}}_{2,1} \left( \kappa ; x_{1}, x_{2} \right) = &
      \frac{1}{\sqrt{2}} \left[ u_{0} \left( \kappa ; x_{1} \right) u_{1} \left(
      \kappa ; x_{2} \right) \right. \nonumber \\[0.2cm]
      & \left. \hspace*{0.5cm} \pm u_{1} \left( \kappa ; x_{1} \right) u_{0} 
      \left( \kappa ; x_{2} \right) \right] \hspace*{0.3cm} , 
      \nonumber \\[0.2cm]
      \Psi^{\textrm{ni}}_{3} \left( \kappa ; x_{1}, x_{2} \right) = & u_{1} 
      \left( \kappa ; x_{1} \right) u_{1} \left( \kappa ; x_{2} \right) 
      \label{eq:two_part_eigen_C00} \hspace*{0.3cm} , 
   \end{align}
with the two-particle eigenenergies given by corresponding combinations of the 
single-particle energies, $E^{\textrm{single}}_{i} \left( \kappa \right)$.  From
Fig. \ref{fig:two_part_eig_spec}(a), one notes that as $\kappa$ is increased all
two-particle eigenstates in the lowest band become degenerate.  This degeneracy
follows automatically from the degeneracy of the states $u_{0} \left( \kappa ; x
\right)$ and $u_{1} \left( \kappa ; x \right)$ seen in the single-particle case 
(see Fig. \ref{fig:sin_part_eig_spec}).  The states, which are symmetric 
(solid lines) and antisymmetric (dashed lines) under exchange, are indicated,
corresponding to boson and fermion pairs.

The effect of introducing interactions between the two bosons is displayed in 
Fig. \ref{fig:two_part_eig_spec}(b) - (d).  In Fig.
\ref{fig:two_part_eig_spec}(b) a scaled interaction coupling of $g_{1D} = 1$ is
considered.  The symmetric states are shifted upwards in energy as a result of 
the repulsive interactions while the antisymmetric states remain unaltered.  In 
the limit of large $\kappa$ one now observes two pairs of degenerate levels, as 
opposed to the set of four degenerate states seen in the non-interacting case.  
The energy separation of these two pairs of levels is monotonically increasing 
with increasing $\kappa$.  Increasing the interaction coupling further, Fig. 
\ref{fig:two_part_eig_spec}(c), leads to one of the symmetric states being 
promoted above the higher-lying antisymmetric state for small $\kappa$, but with 
increased $\kappa$ the normal ordering is restored.  Finally, Fig. 
\ref{fig:two_part_eig_spec}(d) depicts the same spectrum in the limit of strong
repulsion: $g_{1D} = 10$.  The lowest symmetric state now follows closely the 
energy profile of the lowest antisymmetric state.  This feature is a universal 
property for a system of strongly interacting bosonic particles in 1D.  In the 
limit of $g_{1D} \rightarrow \infty$ the bosonic particles become impenetrable, 
and one enters the so-called Tonks-Girardeau regime \cite{ton36,gir60}.  The 
Fermi-Bose mapping \cite{gir60, yuk05} allows, for example, the ground state of 
the bosonic system to be given by
   \begin{equation}
   \label{eq:ground_state_tg}
      \hspace*{-0.3cm} 
      \Psi_{0} \left( x_{1}, x_{2} \right) = \left| \Psi^{\textrm{ni}}_{1} 
      \left( x_{1}, x_{2} \right) \right| \hspace*{0.3cm} .
   \end{equation}
The similarity of the energy of the symmetric ground state and the lowest 
antisymmetric state, seen in Fig. \ref{fig:two_part_eig_spec}(d), is an
indication that the system is approaching this Tonks-Girardeau regime.

\section{\label{sect:ground_state} Ground-state properties}
\begin{figure}[!ht]
      \begin{center}  
         \includegraphics[scale = 0.95]{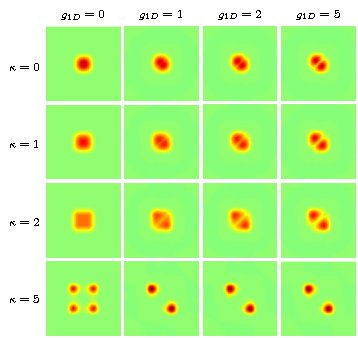} 
      \caption{\label{fig:psi00} Color scale plots of the ground state
       wavefunction, $\Psi_{0} \left( x_{1}, x_{2} \right)$, for a boson pair in
       a double-well potential.  The color scale runs from blue (largest 
       negative value) through to red (largest positive value). The different 
       columns represent different values for the interaction coupling.  The 
       values considered are  $g_{1D} = 0$ (column 1), $g_{1D} = 1$ (column 2), 
       $g_{1D} = 2$ (column 3) and $g_{1D} = 5$ (column 4).  In each case the 
       effect of varying the barrier height is illustrated down a given column. 
       Row 1 corresponds to $\kappa = 0$, row 2 to $\kappa = 1$, row 3 to 
       $\kappa = 2$ and row 4 to $\kappa = 5$.  Results have been obtained from 
       the DVR method with $N = 81$ mesh points in each dimension and a mesh 
       spacing of $h = 0.14$.  Each individual plot spans the range $-5.6 <
       x_{1}, x_{2} < 5.6$.}
       \end{center}
\end{figure}

The ground state wavefunction, $\Psi_{0} \left( x_{1}, x_{2} \right)$, for two 
interacting particles, is presented in Fig. \ref{fig:psi00}.  The individual 
color scale plots will be referenced using standard (\textit{row, column}) matrix
notation.

In the non-interacting case and for $\kappa = 0$, plot (1,1), the wavefunction 
is distributed fairly isotropically about the centre of the trap.  Moving down
this column, increasing $\kappa$, the wavefunction expands slightly in both 
dimensions and takes on a more rectilinear appearance, e.g. plot (3,1).  For 
small values of $\kappa$, the potential resembles that of a square well.  The 
$x^{4}$ term gives rise to a steep boundary and the distribution of the two 
(independent) particles will be quite uniform, leading to the distribution seen
in (3,1).  As the value of $\kappa$ is increased the wavefunction begins to 
segregate into four quadrants with suppression in the region of the barrier (i.e.
along the lines $x_{1} = 0$ and $x_{2} =0$).  This effect is seen, quite
markedly, in plot (4,1).  Considering the energy level spectrum in Fig. 
\ref{fig:two_part_eig_spec}(a) one can see that for $\kappa = 2$ one has not yet 
reached the insulator limit, whereas for $\kappa = 5$ one is deep within this 
insulator regime, for which degeneracy is observed for the four lowest 
two-particle levels.

Turning to the fourth column of Fig. \ref{fig:psi00}, plot (1,4) shows the case 
of no barrier ($\kappa = 0$).  The repulsive interaction precludes any overlap of
the particles.  The effect of increasing the barrier height to $\kappa = 1$,
$\kappa = 2$ and $\kappa = 5$ can be seen in plots (4,2), (4,3) and (4,4), 
respectively.  Again, for small values of $\kappa$, the wavefunction 
distribution expands slightly in ($x_{1}, x_{2}$) space, but now the presence of
repulsive interactions distorts the wavefunction along the line $x_{1} = -
x_{2}$.  In the insulator limit, as we have for (4,4), one sees that the 
wavefunction has split into two clear lobes.

The behavior observed in Fig. \ref{fig:psi00} correlates closely to the 
behavior reported for the $\delta$-split potential in \cite{murA07}: the
segregation of the wavefunction distribution into four quadrants, and the 
vacancy of two of these quadrants owing to the introduction of repulsive 
interactions.  These features are essentially generic for double-well systems.

\subsection{\label{subsect:mom_dist_ground} Momentum distribution}
The reduced single-particle density matrix (RSPDM) has proven to be an 
extremely useful mathematical construct in the analysis of pair correlations 
\cite{col_re}.  For the two-particle system considered here, the RSPDM, 
$\rho_{i} \left( x, x' \right)$, for a given eigenstate, $\Psi_{i} \left( 
x_{1}, x_{2} \right)$, is defined to be
   \begin{equation}
   \label{eq:rspd_def}
      \rho_{i} \left( x, x' \right) = \int_{- \infty}^{+ \infty} \Psi_{i}
      \left( x, x_{2} \right) \Psi_{i} \left( x', x_{2} \right) d x_{2} 
      \hspace*{0.3cm} .
   \end{equation}
This object has been analyzed in detail for the ground state of two particles in 
a $\delta$-split potential \cite{murA07}.  The behavior of $\rho_{0} \left( x,
x' \right)$ for the double-well, presented here, exhibits the same gross features
as have been observed in \cite{murA07} for the $\delta$-split trap problem.  
Instead, in this section we focus on the momentum distributions for this system.

\begin{figure}[!ht]
      \begin{center}  
         \includegraphics[scale = 0.32]{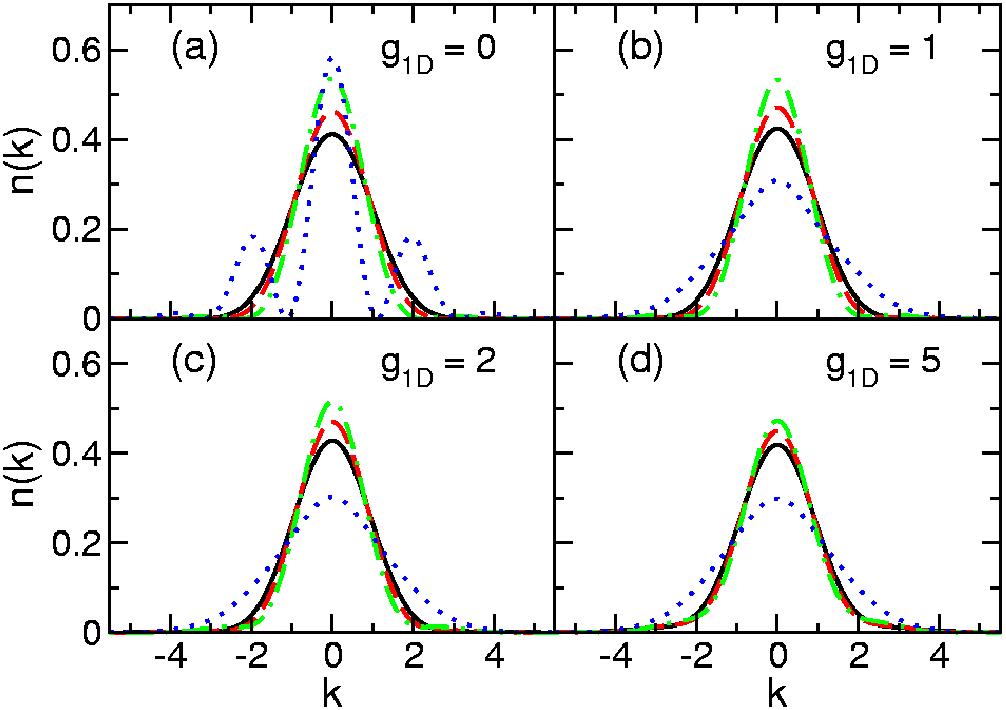} 
      \caption{\label{fig:mom_dist_00} Momentum distribution for the ground 
      state of a system of two bosons confined by a double-well potential.  Four
      different values of interaction strength are considered: (a) $g_{1D} = 0$,
      (b) $g_{1D} = 1$, (c) $g_{1D} = 2$ and $g_{1D} = 5$.  In each figure the
      effect of varying the barrier height is also illustrated.  Barrier heights
      considered are $\kappa = 0$ (solid line), $\kappa = 1$ (dashed line),
      $\kappa = 2$ (dot-dash line) and $\kappa = 5$ (dotted line).}
      \end{center}
   \end{figure}

The reciprocal momentum distribution for the $i^{th}$ eigenstate, $n_{i} \left( 
k \right)$, is calculated  from the corresponding reduced single-particle 
density, $\rho_{i} \left( x, x' \right)$, through Fourier transform
   \begin{equation}
   \label{eq:mom_dist_int}
      n_{i} \left( k \right) \equiv \left( 2 \pi \right)^{-1} \int_{- \infty}^{+
      \infty} \int_{- \infty}^{+ \infty} \rho_{i} \left( x, x' \right)
      \textrm{e}^{- \imath k \left( x - x' \right)} dx dx' \hspace*{0.3cm} ,
   \end{equation}
where $\, \int_{- \infty}^{+ \infty} n_{i} \left( k \right) dk = 1 \,$.  
Equivalently, one may obtain the momentum distribution for this eigenstate by 
considering the diagonalization of $\rho_{i} \left( x, x' \right)$.  
Specifically, the eigenvalue equation is 
   \begin{equation}
   \label{eq:rspdm_diag}
      \int_{-\infty}^{+\infty}\rho_{i} \left( x, x' \right) \phi_{ij} \left( x' 
      \right) dx' = \lambda_{ij} \phi_{ij} \left( x \right) \hspace*{0.3cm} ,
   \end{equation}
where $\lambda_{ij}$, represents the fractional population of the `natural
orbital' $\phi_{ij} \left( x \right)$ such that $\sum_{j} \lambda_{ij} = 1$,
for each $i$.  Using numerical quadrature allows one to rewrite 
(\ref{eq:rspdm_diag}) as a linear equation.  The momentum distribution, $n_{i} 
\left( k \right)$, may then be obtained from the relation
   \begin{equation}
   \label{eq:mom_dist}
      n_{i} \left( k \right) = \sum_{j} \lambda_{ij} \left| \mu_{ij} \left( k
      \right) \right|^{2} \hspace*{0.3cm} ,
   \end{equation}
where $\mu_{ij} \left( k \right)$ denotes the Fourier transform of the natural
orbital $\phi_{ij} \left( x \right)$, 
   \begin{equation}
   \label{eq:ft_nat_orb}
      \mu_{ij} \left( k \right) = \frac{1}{\sqrt{2 \pi}}\int_{-\infty}^{+\infty}
      \phi_{ij} \left( x \right) e^{- \imath k x} dx \hspace*{0.3cm} .
   \end{equation}
The momentum distribution for the ground state is presented in Fig. 
\ref{fig:mom_dist_00}.  The distributions presented correspond to $g_{1D} = 0$ 
(a), $g_{1D} = 1$ (b), $g_{1D} = 2$ (c) and $g_{1D} = 5$ (d).  Also within each
figure, the distributions arising for several different values for the barrier 
height ($\kappa$) are illustrated.
In the non-interacting case, Fig. \ref{fig:mom_dist_00}(a), one observes an
initial peaked distribution for $\kappa = 0$.  Increasing $\kappa$ enhances the
peak and narrows the distribution.  With increasing barrier, the ground-state
wavefunction adapts to spread over the available interval, leading to this
reciprocal narrowing in momentum space.  Further increase in $\kappa$ means that 
the particles begin to experience the effect of the double-well.  As the
particles are non-interacting, the system displays a single-particle behavior.  
For a value of $\kappa = 5$ (insulator regime) the particle splits between the 
wells and the momentum distribution displays prominent second-order peaks, seen 
in Fig. \ref{fig:mom_dist_00}(a).  The momentum distributions can be observed by 
scattering or free expansion of the particles in the absence of a confining 
potential.  From this perspective the second-order peaks correspond to the 
interference fringes that arise from two coherent matter wave sources.

Introducing an interaction encourages localization and has the effect of 
removing these secondary peaks.  Once again, for small values of $\kappa$, e.g. 
$\kappa = 1$ (dashed line) and $\kappa = 2$ (dot-dash line), the momentum 
distribution becomes increasingly peaked and narrower.  In the presence of 
interactions the particles are restricted to separate wells and the interference 
effects are lost.  In addition, the localization of the particles leads to a 
broadening of the momentum distribution, as observed in Fig. 
\ref{fig:mom_dist_00}(b), (c) and (d), for $\kappa = 5$ (dotted line).  It is 
also observed that, in the absence of any barrier, $\kappa = 0$ (solid line), 
one sees the emergence of higher-energy wings for increasing interaction, 
$g_{1D}$.  Similar high-energy wings have been reported in the TG regime for free
space, \cite{ols98}, and harmonic confinement, \cite{gir01}.
  
\subsection{\label{subsect:entropy_ground} Von Neumann entropy}
Entanglement is a fundamental expression of information content and is 
responsible for the increased efficiency of some quantum algorithms over their 
classical counterparts.  Previous authors have shown that the von 
Neumann entropy of the RSPDM is a good measure of entanglement for a 
system of two bosons \cite{pas01, li01, sun06}.  For the case of two 
indistinguishable particles, determination of whether or not the two subsystems 
are entangled requires that one considers both the von Neumann entropy of the 
reduced single-particle density matrix, and the Schmidt number i.e. number of 
non-zero eigenvalues ($\lambda_{ij}$) obtained in the diagonalization of 
$\rho_{i}$, Eq. \eqref{eq:rspdm_diag}, \cite{ghi03, ghiA04, ghiB04}.  We use 
the von Neumann entropy to quantify the entanglement in the position 
coordinates, $x_{1}$ and $x_{2}$, of the particle pair.

Following the diagonalization of the reduced single-particle density, Eq.
(\ref{eq:rspdm_diag}), the von Neumann entropy for the $i^{\textrm{th}}$ 
eigenstate of the two-particle system ($S_{i}$) is obtained from,
 \begin{equation}
 \label{eq:von_neumann}
    S_{i} = -\sum_{j} \lambda_{ij} \log_{2} \lambda_{ij} \hspace*{0.3cm} .
 \end{equation}
 
\subsubsection{\label{subsubsect:entropy_v_interaction_ground} Variation of von 
Neumann entropy with interaction strength}
\begin{figure}[!ht]
      \begin{center}  
         \includegraphics[scale = 0.3]{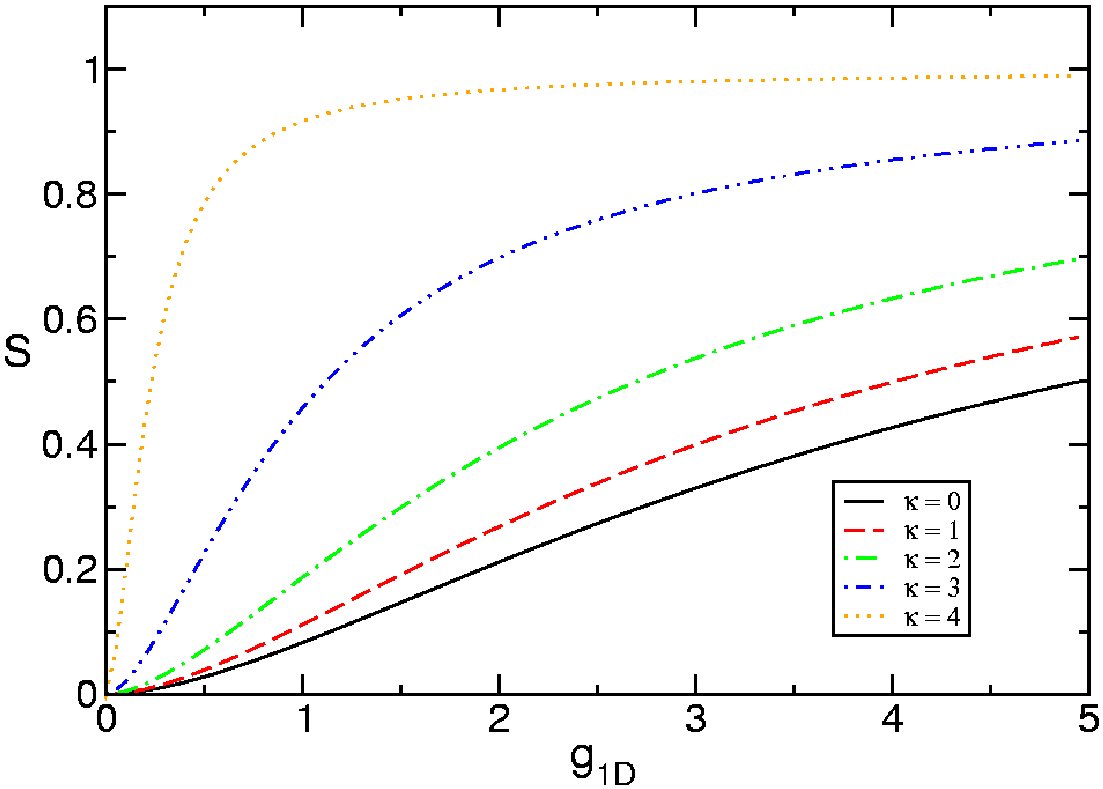} 
      \caption{\label{fig:entropy_v_int_ground} Von Neumann entropy, $S$, as a
      function of the interaction strength, $g_{1D}$.  This dependence is also
      illustrated for a number of different barrier heights: $\kappa = 0$ (solid
      line), $\kappa = 1$ (dashed line), $\kappa = 2$ (dot-dash line), $\kappa =
      3$ (dot-dot-dash line) and $\kappa = 4$ (dotted line).  One observes an 
      increased sensitivity to $g_{1D}$ as the barrier is raised.}
      \end{center}
   \end{figure}

Variation of the von Neumann entropy with $g_{1D}$, for the ground state of this
system, is plotted in Fig. \ref{fig:entropy_v_int_ground}.
Examining the lowest solid line ($\kappa = 0$), when no interactions are present 
($g_{1D} = 0$) then $S = 0$ as one expects.  The product states (with correct
symmetrization) given in Eq. (\ref{eq:two_part_eigen_C00}) represent the
eigenstates of the non-interacting system.  Introducing a small interaction has
the effect of introducing correlations and results in a non-zero entropy. 
Increasing the interaction strength leads to an increasing entropy, saturating at
$S \approx 1$, as for the harmonic potential \cite{murA07, sun06}.  This
behavior can be related to fermionization.  As the repulsive interactions 
increase, the system enters the TG regime.  In this regime the ground state of 
the system can be represented by the corresponding system of two non-interacting
fermions, with correct symmetrization.  In terms of the eigenfunctions 
prescribed in Eq. (\ref{eq:two_part_eigen_C00}), the ground state of the system
is given by $\left| \Psi^{\textrm{ni}}_{1} \left( \kappa ; x_{1}, x_{2} \right) 
\right| = \left| \frac{1}{\sqrt{2}} \left[ u_{0} \left( \kappa ; x_{1} \right) 
u_{1} \left( \kappa ; x_{2} \right) - u_{1} \left( \kappa ; x_{1} \right) u_{0}
\left( \kappa ; x_{2} \right) \right] \right|$.  The antisymmetric state, 
$\Psi_{1}$, in the presence of point-like interactions, will always give $S =
1$.  The ground state of the system becomes degenerate with this antisymmetric 
state in the limit of hard-core interactions.  

Fig. \ref{fig:entropy_v_int_ground} also displays the effect of increasing the 
barrier height, $\kappa$.  As the system tends towards the insulator regime, the
entropy of the system becomes increasingly sensitive to changes in $g_{1D}$, 
about $g_{1D} = 0$.  This effect was also reported in \cite{murA07} for the 
$\delta$-split trap, suggesting that this is another generic feature associated
with double-well potentials.  The increased barrier height reduces tunneling
between the wells.  For any increase in the interaction strength, the 
two-particle wavefunction will attempt to redistribute so as to minimize this 
interaction.  However, with the increased barrier height the wavefunction is 
forced to remain more localized, and is restricted in its redistribution.

\subsubsection{\label{subsubsect:entropy_v_barrier_ground} Variation of von 
Neumann entropy with barrier height}
\begin{figure}[!ht]
      \begin{center}  
         \includegraphics[scale = 0.3]{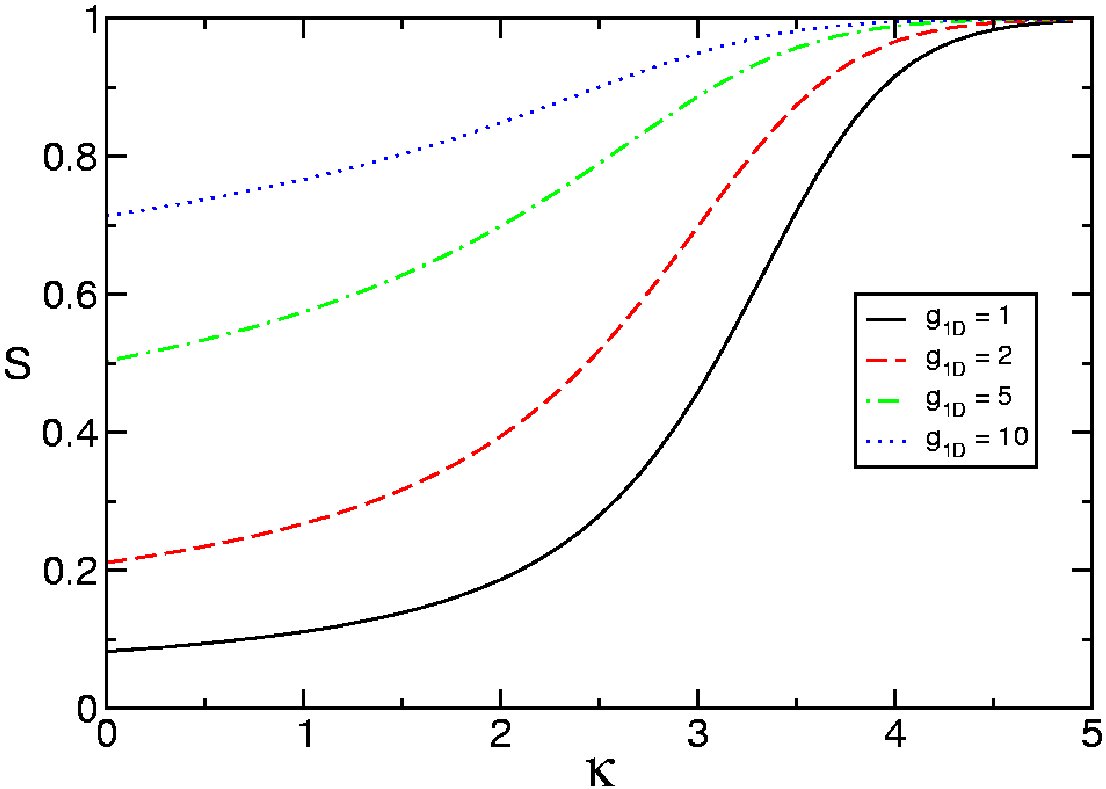} 
      \caption{\label{fig:entropy_v_barrier_ground} Von Neumann entropy, $S$, as
      a function of the barrier height, $\kappa$.  The dependence is examined
      for a number of different interaction coupling strengths: $g_{1D} = 1$ 
      (solid line), $g_{1D} = 2$ (dashed line), $g_{1D} = 5$ (dot-dash line) and 
      $g_{1D} = 10$ (dotted line).  In all cases $S$ saturates at a value of 
      unity in the limit $\kappa \rightarrow \infty$.}
      \end{center}
   \end{figure}

Variation of the von Neumann entropy with $\kappa$, for the ground state of this
system, is plotted in Fig. \ref{fig:entropy_v_barrier_ground}. 
The basic trends bear a striking resemblance to those seen for the
$\delta$-split trap, \cite{murA07}.  Specifically, one observes that the initial
entropy of the system (i.e. for $\kappa = 0$) is dictated by the interaction 
strength of the system, $g_{1D}$.  The larger is $g_{1D}$, the larger is the 
initial value of $S$, as is consistent with Fig. \ref{fig:entropy_v_int_ground}.  
Increasing the height of the barrier then has the effect of increasing the 
entropy of the system towards $S = 1$. In the limit of large barrier heights the
entropy of the system saturates at $S=1$, regardless of the value of interaction
strength (the notable exception being the non-interacting case, for which $S$ is
identically equal to zero for all $\kappa$).  This saturation at $S = 1$ 
corresponds to the loss of entanglement.

When the system enters the insulator limit there is an implicit exchange 
uncertainty in the state of the system, arising from the indistinguishable nature
of the particles.  As such, these correlations cannot be exploited in any 
meaningful quantum information protocol and the system is regarded as
non-entangled.  This diagnosis also follows from the criteria set out in 
\cite{ghi03, ghiA04, ghiB04} as, in the limit $\kappa \rightarrow \infty$ then 
$S \rightarrow 1$ and the Schmidt number can be seen to approach a value of 2 
(not shown here).  By the criteria outlined in \cite{ghiA04}, any state for 
which $S = 1$  and with a Schmidt number of 2 must be regarded as non-entangled.

In contrast to the $\delta$-split trap, the entropy dependence $S \left( \kappa 
\right)$ for the double-well system is quite sigmoidal.  The separation of wells 
only becomes apparent for large values of $\kappa$ (i.e. $\kappa > 3$).  From
Fig. \ref{fig:entropy_v_barrier_ground} one can identify $2 < \kappa < 3$ as the 
interval over which the entropy makes its most rapid variation.

\subsubsection{\label{subsubsect:bose_hubbard_entropy} Von Neumann entropy in 
the Bose-Hubbard model} 
One may examine the von Neumann entropy of the ground state within the formalism
of the Bose-Hubbard model presented in Eq. \eqref{eq:bose_hubbard_ham}.  Using a
Fock basis for the two-particle system of the form $ \left. \left|
n_{L} n_{R} \right. \right>$, where $n_{L(R)}$ represents the number of 
particles in the left (right) well, leads to three basis states : $ \left. \left|
2 0 \right. \right>$, $ \left. \left| 1 1 \right. \right>$ and $ \left. \left| 0
2 \right.  \right>$.  

Thus, in terms of this basis the Hamiltonian \eqref{eq:bose_hubbard_ham} may be 
written in matrix form as
   \begin{equation}
   \label{eq:bose_hubbard_ham_matrix}
      \hat{H} = 
      \left(
         \begin{array}{ccc}
            2 \epsilon + 2 U & \sqrt{2} J & 0 \\[0.2cm]
	    \sqrt{2} J & 2 \epsilon & \sqrt{2} J \\[0.2cm]
	    0 & \sqrt{2} J & 2 \epsilon + 2 U    
         \end{array}
      \right) \hspace*{0.3cm} .
   \end{equation}
with eigenvalues of this Hamiltonian follow from some simple algebra
 \begin{align}
    E_{-} = & 2 \epsilon + U - \sqrt{U^{2} + 4 J^{2}} \nonumber \\[0.2cm]
    E_{\textrm{mid}} = & 2 \epsilon + 2 U \nonumber \\[0.2cm]
    E_{+} = & 2 \epsilon + U + \sqrt{U^{2} + 4 J^{2}}
    \label{eq:bose_hubbard_eig_val} \hspace*{0.3cm} .
 \end{align}
The eigenvectors corresponding to the eigenvalues presented in Eq.
\eqref{eq:bose_hubbard_eig_val} are found to be
   \begin{equation}
   \label{eq:bose_hubbard_eig_vec}
      \left. \left| \Psi_{\textrm{mid}} \right. \right> = \frac{1}{\sqrt{2}}
      \left(
      \begin{array}{c}
         1 \\
         0 \\
         -1
      \end{array}
      \right) \hspace*{0.5cm}
      \left. \left| \Psi_{\pm} \right. \right> = \mathcal{N}_{\pm} 
      \left(
      \begin{array}{c}
         1 \\
         \frac{2 \sqrt{2} J}{U \pm \sqrt{U^{2} + 4 J^{2}}} \\
         1
      \end{array}
      \right) \hspace*{0.3cm} ,
   \end{equation}
where $\mathcal{N}_{\pm}$ represent normalization factors and $\left. \left|
\Psi_{\textrm{mid}} \right. \right>$ has odd inversion symmetry.

\begin{figure}[!ht]
      \begin{center}  
         \includegraphics[scale = 0.3]{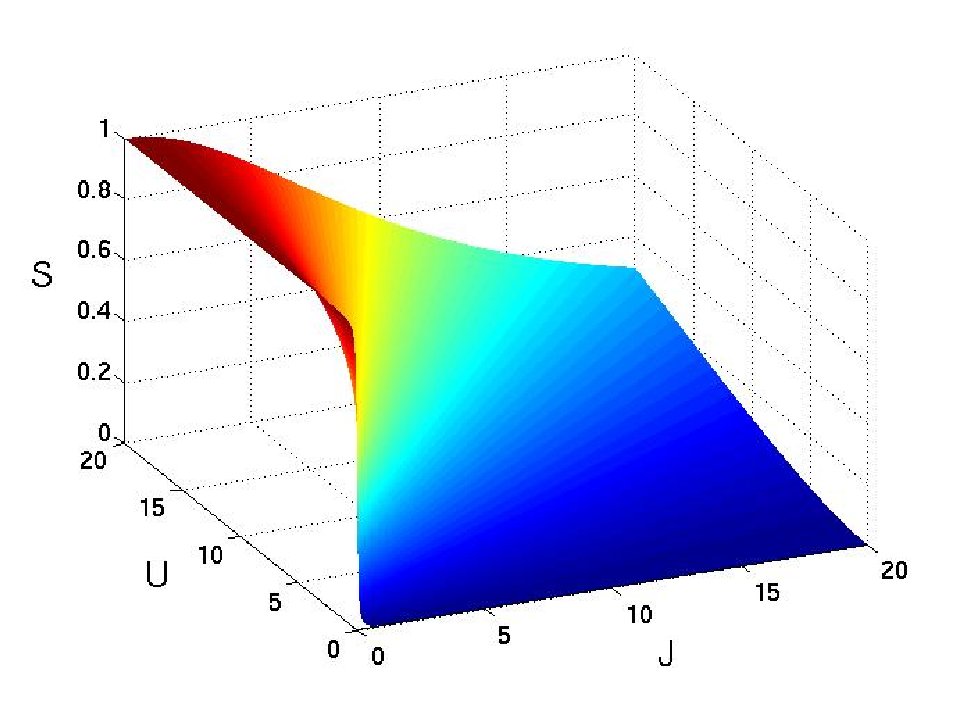} 
      \caption{\label{fig:bhm_entropy_surf_JU} Von Neumann entropy of the ground
      state as a function of the Hamiltonian parameters $J$ and $U$.  The limit
      $J/U \rightarrow 0$ corresponds to the insulator limit and $U/J
      \rightarrow 0$ to the conductor limit.}
      \end{center}
   \end{figure}

Given the two-particle ground state in the Fock basis, $\left. \left| \Psi_{-}
\right. \right>$ , one may determine the reduced single-particle density matrix 
by tracing over the degrees of freedom of either particle.  The single-particle 
basis states may be represented in the form $\left. \left| n_{L}, n_{R} \right. 
\right>$ as $\left. \left| 1 0 \right. \right>$ and $\left. \left| 0 1 \right. 
\right>$.  In turn, the two-particle basis states can be written in the symmetric
form:
   \begin{align}
      \left. \left| 2 0 \right. \right> = & \left. \left| 1 0 \right. 
      \right>_{1} \otimes \left. \left| 1 0 \right. \right>_{2} 
      \nonumber \\[0.2cm]
      \left. \left| 1 1 \right. \right> = & \frac{1}{\sqrt{2}} \left[ \left. 
      \left| 1 0 \right. \right>_{1} \otimes \left. \left| 0 1 \right. 
      \right>_{2}
      + \left. \left| 0 1 \right. \right>_{1} \otimes \left. \left| 1 0 \right. 
      \right>_{2} \right] \nonumber \\[0.2cm]
      \left. \left| 0 2 \right. \right> = & \left. \left| 0 1 \right. 
      \right>_{1}
      \otimes \left. \left| 0 1 \right. \right>_{2}
      \label{eq:two_particle_basis_reexpressed} \hspace*{0.3cm} .
   \end{align}
The eigenvalues of this RSPDM are found to be 
   \begin{align}
      \lambda_{1} = & \mathcal{N}_{-}^{2} \left[ 1 + \frac{4 J^{2}}{\left( U -
      \sqrt{U^{2} + 4 J^{2}} \right)^{2}} - \frac{4 J}{U - \sqrt{U^{2} + 4
      J^{2}}}\right] \nonumber \\[0.2cm]
      \lambda_{2} = & \mathcal{N}_{-}^{2} \left[ 1 + \frac{4 J^{2}}{\left( U -
      \sqrt{U^{2} + 4 J^{2}} \right)^{2}} + \frac{4 J}{U - \sqrt{U^{2} + 4
      J^{2}}}\right] \label{eq:eigval_rspdm_bhm} \hspace*{0.3cm} .
   \end{align}
The variation of the ground-state entropy ($S$) with the model parameters $J$ 
and $U$ is depicted as a surface plot in Fig. \ref{fig:bhm_entropy_surf_JU}.
It is noted that the qualitative behavior of the entropy displayed in Figs. 
\ref{fig:entropy_v_int_ground} and \ref{fig:entropy_v_barrier_ground} for 
varying $g_{1D}$ and $\kappa$, respectively, is reflected in the Bose-Hubbard 
model with variation of the parameters $U$ and $J$.  This tight-binding
approximation is poor in the limit $U/J \rightarrow 0$, but is an accurate
representation in the insulator limit.

\section{\label{sect:excited_states} Excited states}
Attention is now turned to the three lowest excited states which, together
with the ground state, represent the lowest energy band of the two-particle 
double-well system.  In the non-interacting case, with spectrum depicted in Fig.
\ref{fig:two_part_eig_spec}(a), the three lowest excited states may be 
represented as given in Eq. \eqref{eq:two_part_eigen_C00}, with one of these 
states being antisymmetric and two of them symmetric.  Variation of parameters 
$\kappa$ and $g_{1D}$ can lead to reordering of the energy eigenvalues, as 
observed in Fig. \ref{fig:two_part_eig_spec}.  However, in this section, the 
study of the excited states of this two-particle system will be restricted to 
these three states of the lowest band, identifiable through their symmetry.  
Henceforth the term `first-excited state' refers to the lowest energy 
antisymmetric state ($\Psi_{1}$), `second-excited state' refers to the 
second-lowest energy symmetric state ($\Psi_{2}$) and `third-excited state' 
refers to the third-lowest lying symmetric state ($\Psi_{3}$).
 
\subsection{\label{subsect:wavefunction_excited} Two-particle excitations}
\begin{figure}[!ht]
      \begin{center}  
         \includegraphics[scale = 0.95]{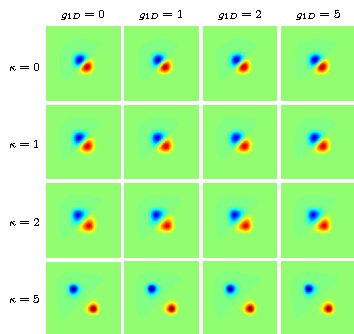} 
      \caption{\label{fig:psi01} Color scale plots of the first-excited
       eigenfunction, $\Psi_{1} \left( x_{1}, x_{2} \right)$, for a pair of 
       particles in a double-well potential.  The color scale runs from blue
       (largest negative value) to red (largest positive value).  The different 
       columns represent different values for the interaction coupling.  The 
       values considered are  $g_{1D} = 0$ (column 1), $g_{1D} = 1$ (column 2), 
       $g_{1D} = 2$ (column 3) and $g_{1D} = 5$ (column 4).  In each case the 
       effect of varying the barrier height is illustrated down a given column. 
       Row 1 corresponds to $\kappa = 0$, row 2 to $\kappa = 1$, row 3 to 
       $\kappa = 2$ and row 4 to $\kappa = 5$.  Notice how the antisymmetric 
       state is completely independent of the interaction parameter, $g_{1D}$.  
       Results have been obtained using the same DVR method described for 
       Fig. \ref{fig:psi00}.} 
       \end{center}
\end{figure}

\begin{figure}[!ht]
      \begin{center}  
         \includegraphics[scale = 0.95]{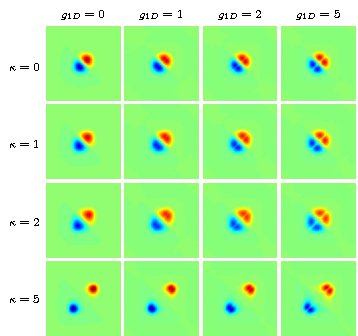} 
      \caption{\label{fig:psi02} Color scale plots of the second-excited
       eigenfunction, $\Psi_{2} \left( x_{1}, x_{2} \right)$, for a pair of
       particles in a double-well potential.  The color scale runs from blue
       (largest negative value) to red (largest positive value).  The different 
       columns represent different values for the interaction coupling and the 
       effect of varying the barrier height is illustrated down a given column. 
       The same values for the parameters $g_{1D}$ and $\kappa$, as considered in
       Fig. \ref{fig:psi01}, are examined here.  Results have been obtained 
       using the same DVR method described for Fig. \ref{fig:psi00}.}
       \end{center}
\end{figure}

\begin{figure}[!ht]
      \begin{center}  
         \includegraphics[scale = 0.95]{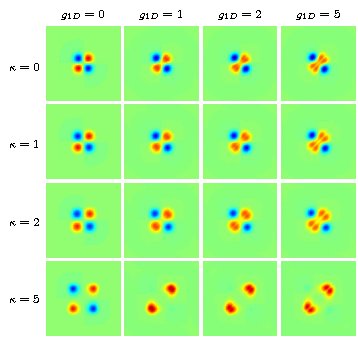} 
       \caption{\label{fig:psi03} Color scale plots of the third-excited
       eigenfunction, $\Psi_{3} \left( x_{1}, x_{2} \right)$, for a pair of
       particles in a double-well potential.  The color scale runs from blue
       (largest negative value) to red (largest positive value).  The different 
       columns represent different values for the interaction coupling and the 
       effect of varying the barrier height is illustrated down a given column. 
       The same values for the parameters $g_{1D}$ and $\kappa$, as considered in
       Fig. \ref{fig:psi01}, are examined here.  Results have been obtained 
       using the same DVR method described for Fig. \ref{fig:psi00}.} 
       \end{center}
\end{figure}

The wavefunctions for $\Psi_{1,2,3}$ are represented, by means of color scale
plots, in Figs. \ref{fig:psi01}, \ref{fig:psi02} and \ref{fig:psi03}, 
respectively.  Again, the standard (\textit{row, column}) notation is used
to reference individual subplots.  The color scale is consistent across all 
wavefunction plots, permitting direct comparison between Figs. \ref{fig:psi00},
\ref{fig:psi01}, \ref{fig:psi02} and \ref{fig:psi03}.

Fig. \ref{fig:psi01} represents the ground state for a system of two 
spin-aligned fermions ($\Psi_{1}$), which is identically zero along the line 
$x_{1} = x_{2}$ and, thereby, unaffected by the zero-ranged interaction.  
Considering Fig. \ref{fig:psi01}, moving along a given row (i.e. increasing
repulsion for a fixed barrier), the wavefunction plots remain unchanged, 
illustrating the independence of this state with respect to interaction
strength, $g_{1D}$.  As $\kappa$ increases (i.e. down any column) the positive
and negative lobes along the $x_{1} = -x_{2}$ diagonal become more widely 
separated indicating isolation into separate wells.  Once again the wavefunction
density in these two quadrants correspond to the situation where particle 1 is 
in the left well ($x_{1} < 0$) and particle 2 is in the right ($x_{2}> 0$), and
vice versa.  In the limit of a large barrier, the ground state becomes 
degenerate with this antisymmetric state.  The wavefunction plots are almost 
identical (except for sign) for $\kappa = 5$ as seen, for example, by comparing
Fig. \ref{fig:psi00} (4,4) and Fig. \ref{fig:psi01} (4,4).  Furthermore, as 
already discussed, one expects the ground state to become degenerate with this
antisymmetric state in the limit of $g_{1D} \rightarrow \infty$, for all
$\kappa$.  This degeneracy is evidenced by comparing the fourth column in Fig. 
\ref{fig:psi00} to any column in Fig. \ref{fig:psi01}.  Even at this finite 
interaction strength ($g_{1D} = 5$) the equivalence of these two states is 
apparent.  Finally, from each of the plots in Fig. \ref{fig:psi01} it is clear
that this eigenstate is of odd parity, such that $\Psi_{1} \left( x_{1}, x_{2}
\right) = - \Psi \left( -x_{1}, -x_{2}  \right)$.  

Fig. \ref{fig:psi02} depicts the second-excited state for the system of two
bosons in a double-well potential and, as with $\Psi_{1}$, this state exhibits 
odd parity: $\Psi_{2} \left( x_{1}, x_{2} \right) = - \Psi_{2} \left( -x_{1},
-x_{2} \right)$.  The case of no barrier ($\kappa = 0$) and no interaction 
($g_{1D} = 0$) is illustrated in Fig. \ref{fig:psi02} (1,1).  The eigenstate is 
composed of two lobes which correspond to both particles co-existing on the same 
side of the well.  In the case of no interactions ($g_{1D} = 0$), illustrated in 
column 1, this symmetric eigenstate is degenerate with the antisymmetric state 
considered in Fig. \ref{fig:psi01}.  Repulsive interactions will tend to exclude 
the wavefunction from the line $x_{1} = x_{2}$ (e.g. compare (1,1) to (1,3) or 
(1,4)).  In the Tonks limit this splits each of the upper right and lower left 
lobes.  Considering the effect of the barrier in column 3 ($g_{1D} = 2$), the 
initial wavefunction demonstrates the double-lobe structure.  As the barrier is 
increased to $\kappa = 1$, (2,3), and then $\kappa = 2$, (3,3), the wavefunction 
spreads out in ($x_{1}, x_{2}$) space.  Further increase of the barrier height 
causes the system to move into the insulator limit, plot (4,3), forming two 
isolated lobes in the upper-right and lower-left quadrants.  The eigenstate, in 
this case, corresponds to the physical situation of both particles residing in
either the left well or the right well. 

Finally, the third-excited state is illustrated in Fig. \ref{fig:psi03}. 
In contrast to $\Psi_{1}$ and $\Psi_{2}$, this eigenstate is of even parity such
that $\Psi_{3} \left( x_{1}, x_{2} \right) = \Psi_{3} \left( -x_{1}, -x_{2}
\right)$.   
Scanning down column 1: as the system moves into the insulator limit, the 
eigenstate is composed of four equally-weighted lobes in the four quadrants, 
equivalent to the corresponding ground-state eigenfunction, seen in Fig.
\ref{fig:psi00} (4,1).  In fact from Fig. \ref{fig:two_part_eig_spec}(a), in the 
non-interacting case, the four lowest eigenstates all become degenerate in the 
insulator limit ($\kappa \rightarrow \infty$).  As a consequence, the 
eigenfunction plot (4,1) in Figs. \ref{fig:psi00}, \ref{fig:psi01}, 
\ref{fig:psi02} and \ref{fig:psi03} relate to four degenerate states.

This symmetric eigenstate is non-zero along the line $x_{1} = x_{2}$.  As one 
increases $g_{1D}$ one again observes the exclusion of the wavefunction from this
line (e.g. examining row 1 in Fig. \ref{fig:psi03}).  As barrier height is
increased the wavefunction expands in ($x_{1}, x_{2}$) space and there is some 
suppression of the wavefunction in the region of the rising barrier (i.e. $x_{1}
= 0$ and $x_{2} = 0$).  In the insulator limit, e.g. in plot (4,4) for which 
$\kappa = 5$, the wavefunction in the off-diagonal quadrants vanishes and one 
observes two double-lobes in the lower-left and upper-right quadrants, 
representing the physical situation where both particles reside in the same well.
The degeneracy of $\Psi_{2}$ and $\Psi_{3}$, in the limit $\kappa \rightarrow 
\infty$ (seen in Fig. \ref{fig:two_part_eig_spec}) is manifested in the 
corresponding wavefunction plots.  This is demonstrated by comparing 
corresponding plots in the bottom rows ($\kappa = 5$) of Figs. \ref{fig:psi02} 
and \ref{fig:psi03}.

In the limit of large $\kappa$ (and for any positive interaction), the ground and 
first-excited states correspond to the two particles in separate wells.  By 
contrast, the second- and third-excited states, in the same limit, correspond 
to two particles in the same well.  It follows that an increase in the repulsive
interaction coupling will cause this second pair of levels to be shifted upwards
in energy.  In this way, in the $\kappa \rightarrow \infty$ limit, one observes 
the separation of these two pairs of levels to increase as $g_{1D}$ is increased
(see Fig. \ref{fig:two_part_eig_spec}).  The increasing separation of these 
levels with increasing $\kappa$ is corollary to this.  As $\kappa$ is increased 
the particles become more tightly confined to the individual wells.  This 
increased confinement, for the upper pair of levels, will give rise to an 
increased interaction of the two particles and a subsequent increase in the 
energy of these eigenstates, relative to the lower pair.
 
\subsection{\label{subsect:mom_dist_excited} Momentum distribution}   
   \begin{figure}[!ht]
      \begin{center}  
         \includegraphics[scale = 0.3]{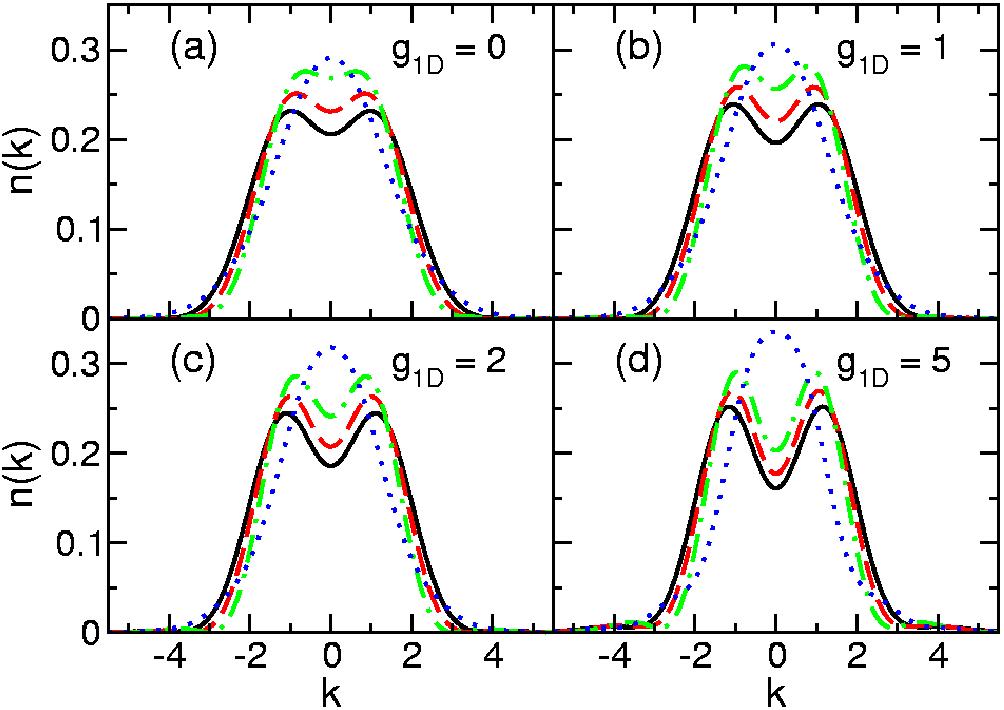} 
      \caption{\label{fig:mom_dist_02} Momentum distribution for the 
      second-excited state ($\Psi_{2}^{\textrm{o}}$) for a system of two atoms 
      confined by the double-well potential $V \left( x \right) = \left( x^{4} -
      \kappa x^{2} \right)$.  Four different values of interaction strength are 
      considered: (a) $g_{1D} = 0$, (b) $g_{1D} = 1$, (c) $g_{1D} = 2$ and 
      $g_{1D} = 5$.  In each figure the effect of varying the barrier height is 
      also illustrated.  Barrier heights considered are $\kappa = 0$ (solid 
      line), $\kappa = 1$ (dashed line), $\kappa = 2$ (dot-dash line) and 
      $\kappa = 5$ (dotted line).  The momentum distribution is identical to 
      that of $\Psi_{1}^{\textrm{o}}$ in the non-interacting limit.  Increasing
      the interaction strength ($g_{1D}$) leads to an increasingly peaked
      distribution.}
      \end{center}
   \end{figure}
The momentum distributions for the excited states are calculated as outlined in
Sec. \ref{subsect:mom_dist_ground}.  The calculated distributions for the 
second-excited state ($\Psi_{2}^{\textrm{o}}$, where the superscript `o' 
indicates the `odd' inversion symmetry of this eigenstate) are displayed in Fig. 
\ref{fig:mom_dist_02}.  For $\kappa = 0$ one observes a double-humped 
distribution that becomes narrower with increasing $\kappa$ and, in the insulator
limit, gives way to a single-peak distribution with high-energy tails.  This is
similar to the result for $\Psi_{1}^{\textrm{o}}$ (not shown).  An increase in 
the interaction coupling has the effect of narrowing the momentum distribution. 
Fig. \ref{fig:psi02} illustrates that increasing $g_{1D}$ will expand the 
wavefunction in ($x_{1}, x_{2}$) space, leading to this reciprocal narrowing in 
momentum space.  At the same time the increased interaction leads to an 
accentuation of the double-peaked structure, observed for small $\kappa$.  
   \begin{figure}[!ht]
      \begin{center}  
         \includegraphics[scale = 0.3]{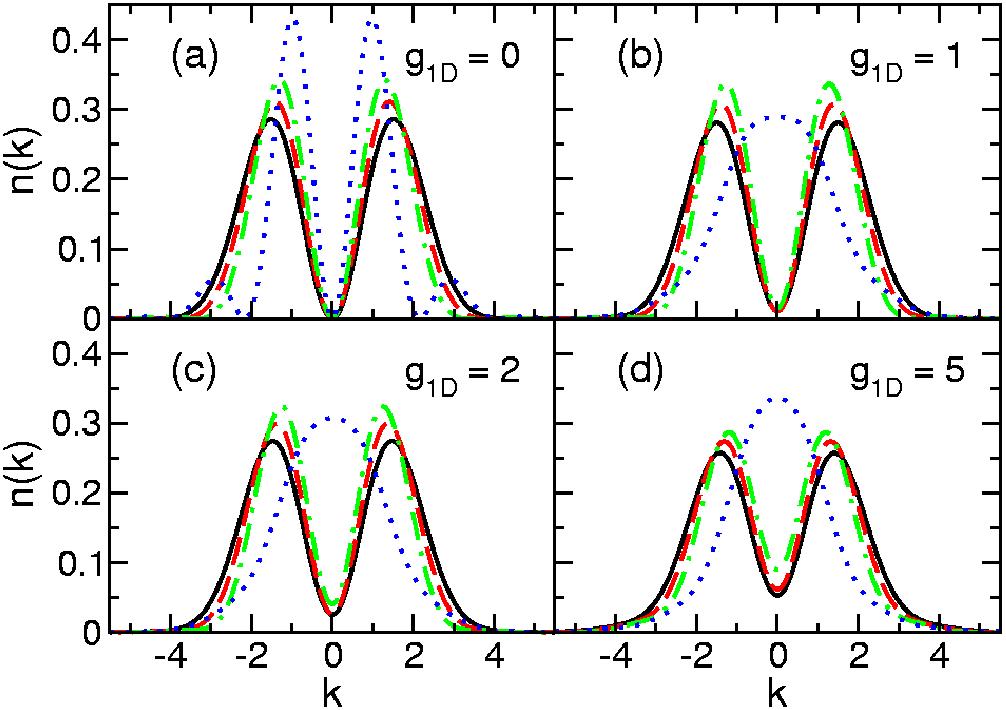} 
      \caption{\label{fig:mom_dist_03} Momentum distribution for the
      third-excited state ($\Psi_{3}^{\textrm{e}}$) for a system of two atoms 
      confined by the double-well potential $V \left( x \right) = \left( x^{4} -
      \kappa x^{2} \right)$.  Four different values of interaction strength are 
      considered: (a) $g_{1D} = 0$, (b) $g_{1D} = 1$, (c) $g_{1D} = 2$ and 
      $g_{1D} = 5$.  In each figure the effect of varying the barrier height is
      also illustrated.  Barrier heights considered are $\kappa = 0$ (solid
      line), $\kappa = 1$ (dashed line), $\kappa = 2$ (dot-dash line) and 
      $\kappa = 5$ (dotted line).  The separable nature of
      $\Psi_{3}^{\textrm{e}}$ in the non-interacting limit, as seen in Eq.
      \eqref{eq:two_part_eigen_C00}, leads to node in momentum distribution for 
      $k = 0$.  Introduction of interactions means that $\Psi_{3}^{\textrm{e}}$
      is no longer separable and node is no longer enforced.}
      \end{center}
   \end{figure}

Fig. \ref{fig:mom_dist_03} illustrates the momentum distribution for the
even-parity state $\Psi_{3}^{\textrm{e}}$.  For the non-interacting case, $g_{1D}
= 0$ (a), a double-mode distribution arises with a node at $k = 0$.  This node is
accounted for due to the separable nature of $\Psi_{3}^{\textrm{e}}$ in the 
non-interacting limit: $\Psi^{\textrm{ni}}_{3} \left( \kappa ; x_{1}, x_{2} 
\right) = u_{1} \left( \kappa ; x_{1} \right) u_{1} \left( \kappa ; x_{2}
\right)$.  Considerable narrowing of this distribution is noted as $\kappa$
is increased and in the insulator limit a second pair of smaller peaks emerges. 
This second pair of peaks may be viewed as interference fringes from each 
particle being distributed between the two wells - compare Fig. \ref{fig:psi00} 
(4,1) and Fig. \ref{fig:psi03} (4,1).  

For increased interaction strength ($g_{1D}$) one continues to observe the 
narrowing of the distribution with increased barrier height.  However, the 
presence of the interactions causes the node at $k=0$ to be removed, as one can 
no longer write the eigenfunction in the separable form given in Eq.
\eqref{eq:two_part_eigen_C00}.  Instead one just observes a strong depression of
the distribution about $k = 0$.  At the same time, the introduction of the 
interactions has the effect of completely removing the double-peaked structure in 
the insulator limit, as is observed for the dotted line ($\kappa = 5$) in each of 
Fig. \ref{fig:mom_dist_03}(b), (c) and (d).  As seen in Fig. \ref{fig:psi03}, in 
the presence of a finite interaction the wavefunction in the off-diagonal 
quadrants vanishes in the insulator limit, and this eigenstate describes a 
situation where both particles occupy one side of the double-well.

\subsection{\label{subsect:entropy_excited} Von Neumann entropy}
As for the ground state, one may obtain the von Neumann entropy for the excited
states of the two-particle system via diagonalization of the reduced
single-particle density matrix.  In this section the dependence of the von 
Neumann entropy, $S$, of the four lowest two-particle states, on the interaction
strength ($g_{1D}$) and the barrier height ($\kappa$) is considered.

\subsubsection{\label{subsubsect:entropy_v_interaction_excited} Variation of 
von Neumann entropy with interaction strength}
   \begin{figure*}[!ht]
      \begin{center}  
         \includegraphics[scale = 0.5]{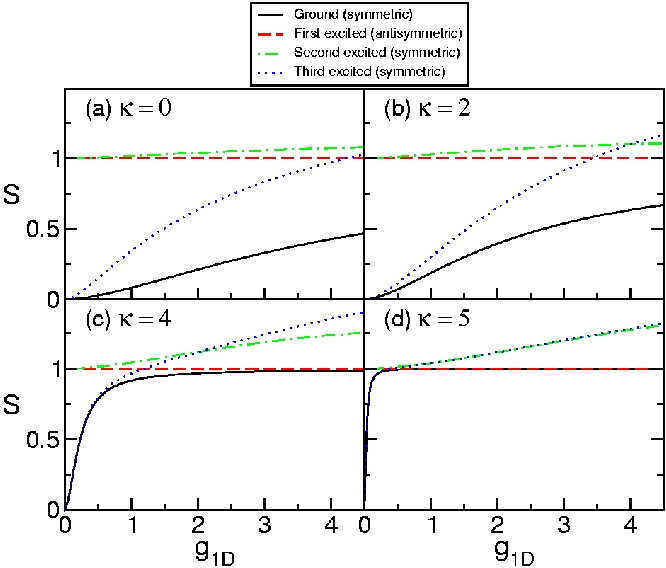} 
      \caption{\label{fig:entropy_v_int_excited} Von Neumann entropy, $S$, as a
      function of the interaction strength ($g_{1D}$) for the four lowest
      two-particle states: $\Psi_{0}^{\textrm{e}}$ (solid line),
      $\Psi_{1}^{\textrm{o}}$ (dashed line), $\Psi_{2}^{\textrm{o}}$ (dot-dash 
      line) and $\Psi_{3}^{\textrm{e}}$ (dotted line).  The dependence is also 
      illustrated for a number of different barrier heights: $\kappa = 0$ (a), 
      $\kappa = 2$ (b), $\kappa = 4$ (c) and $\kappa = 5$ (d).}
      \end{center}
   \end{figure*}

Fig. \ref{fig:entropy_v_int_excited} illustrates the dependence of $S$ on the 
interaction coupling, $g_{1D}$ ($> 0$).  The dependence is examined for four 
different values of barrier height: $\kappa = 0$ (a), $\kappa = 2$ (b), $\kappa
= 4$ (c) and $\kappa = 5$ (d).  For each value of the barrier height the entropy
of the four lowest eigenstates is depicted: ground state (solid line), 
first-excited state (dashed line), second-excited state (dot-dash line) and 
third-excited state (dotted line).  The dependence of the ground-state entropy 
on $g_{1D}$ has already been examined in Fig. \ref{fig:entropy_v_int_ground}, 
however it is useful to replicate these plots here to help inform the 
examination of the excited-state plots.

Several important features are noted.  In all cases the first-excited state 
(dashed line) shows no dependence on the interaction strength, as is expected 
owing to the symmetry of this eigenstate.  Instead, this eigenstate exhibits a 
value of $S = 1$ for all $g_{1D}$.  This value follows from the analytic form 
for this eigenstate, $\Psi^{\textrm{ni}}_{1}$, given by Eq.
\eqref{eq:two_part_eigen_C00}, which holds for all values of $g_{1D}$.  At the
same time, the analytic representations for the three remaining eigenstates are
also given in Eq. \eqref{eq:two_part_eigen_C00}, for $g_{1D} = 0$.  From these
representations it is clear that, in the non-interacting limit, the entropy for 
the ground state (solid line) and third-excited state (dotted line) is always 
zero, as these states may always be represented as direct-product states for 
$g_{1D} = 0$.  In a similar way, the second-excited state (dot-dash line) 
always assumes a value of $S = 1$ in the non-interacting limit.  Once again, 
this may be attributed to the symmetrized form for this state as given by 
$\Psi^{\textrm{ni}}_{2}$ in Eq. \eqref{eq:two_part_eigen_C00}.

Considering the case of $\kappa = 0$, Fig. \ref{fig:entropy_v_int_excited}(a), 
the ground state begins at $S = 0$ and increases monotonically with $g_{1D}$.  
As $g_{1D} \rightarrow \infty$ one enters the TG regime and this ground state
(solid line) becomes degenerate with the first-excited (dashed line) state and 
$S \approx 1$.  By contrast, the second-excited state begins with $S = 1$, as 
discussed, and increases with increasing $g_{1D}$, but at a much slower rate 
than that exhibited by the ground state.  The third excited state (dotted line) 
begins, like the ground state, with $S = 0$ and increases rapidly with 
increasing interaction strength.

Increasing the height of the barrier to $\kappa = 2$, Fig.
\ref{fig:entropy_v_int_excited}(b), one observes qualitatively similar
behavior from all four states except that each state exhibits a more marked
variation in $S$ over the range of $g_{1D}$ examined.  As one moves into the
insulator regime, e.g. $\kappa = 4$, Fig. \ref{fig:entropy_v_int_excited}(c),
the behavior changes quite significantly.  As discussed previously, the ground
state exhibits a very drastic variation with $g_{1D}$, converging very rapidly 
to $S \approx 1$.  The second-excited state still exhibits the same basic 
behavior as noted for smaller $\kappa$ but, once again, the increased barrier 
height leads to an increased sensitivity of this state to variation in
$g_{1D}$.  The third-excited state shows a distinct change in behavior for this 
increased barrier height.  At small values of interaction coupling ($g_{1D} < 1$)
the entropy of this state follows closely that of the ground state.  As 
interaction strength is increased beyond this value then the ground-state entropy
begins to plateau at $S \approx 1$, whilst that of third-excited state continues 
to increase.  Increasing the barrier height to $\kappa = 5$, Fig. 
\ref{fig:entropy_v_int_excited}(d), moves the system deeper into the insulator 
limit and the behavior demonstrated in (c) becomes even more striking.  In 
this case the behavior of the ground-state entropy is more dramatic, with 
the entropy saturating at $S \approx 1$, already, for $g_{1D} \approx 0.5$.  
Again the entropy of the third-excited state follows this trend identically. 
However, where the entropy of the ground state plateaus at $S \approx 1$, the 
entropy of the third-excited state continues to increase and follows now, almost 
identically, the entropy of the second-excited state.  A handle on this 
behavior is provided by the wavefunction plots of Figs. \ref{fig:psi00}, 
\ref{fig:psi02} and \ref{fig:psi03}.  One observes that, in this insulator limit
($\kappa = 5$),  the third-excited state, for small $g_{1D}$, as seen in Fig. 
\ref{fig:psi03} (4,1), closely resembles the ground state in Fig. 
\ref{fig:psi00} (4,1).  For larger interaction couplings ($g_{1D} \ge 1$) the 
eigenfunction for this third-excited state, as seen in Fig. \ref{fig:psi03} 
(4,2) - (4,4), closely resembles that of the second-excited state in Fig. 
\ref{fig:psi02} (4,2) - (4,4). 

\subsubsection{\label{subsubsect:entropy_v_barrier_excited} Variation of von 
Neumann entropy with barrier height}
   \begin{figure*}[!ht]
      \begin{center}  
         \includegraphics[scale = 0.5]{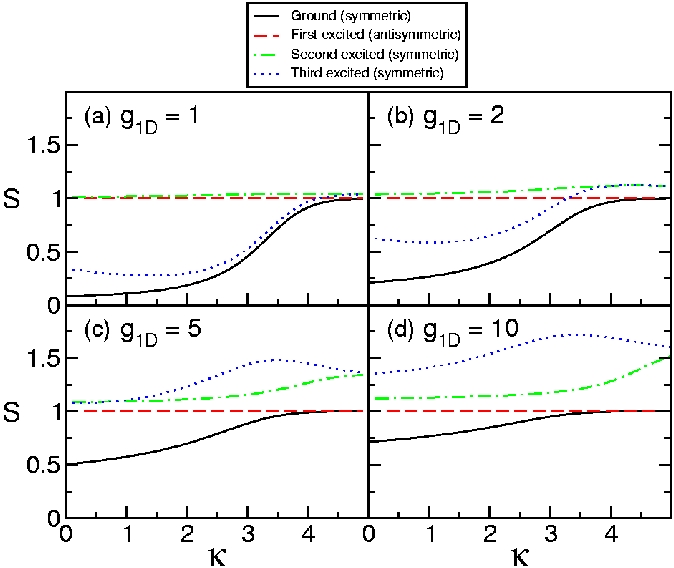} 
      \caption{\label{fig:entropy_v_barrier_excited} Von Neumann entropy, $S$, 
      as a function of the barrier height ($\kappa$) for the four lowest 
      two-particle states: $\Psi_{0}^{\textrm{e}}$ (solid line), 
      $\Psi_{1}^{\textrm{o}}$ (dashed line), $\Psi_{2}^{\textrm{o}}$ (dot-dash 
      line) and $\Psi_{3}^{\textrm{e}}$ (dotted line).  The dependence is 
      examined for a number of different interaction coupling strengths: (a) 
      $g_{1D} = 1$, (b) $g_{1D} =  2$, (c) $g_{1D} = 5$ and (d) $g_{1D} = 10$.}
      \end{center}
   \end{figure*}
   
The variation of von Neumann entropy with barrier height is illustrated in Fig.
\ref{fig:entropy_v_barrier_excited}, for the same four, lowest-energy 
two-particle states.  In this case, four different 
values of interaction coupling are presented: $g_{1D} = 1$ (a), $g_{1D} = 2$
(b), $g_{1D} = 5$ (c) and $g_{1D} = 10$ (d).  Again, in each plot the eigenstates
are represented by the same line types used in Fig.
\ref{fig:entropy_v_int_excited}.

Some general features and behaviors can be noted from these plots.  Again, the 
first-excited state is observed to have an entropy of unity for all $g_{1D}$ and
$\kappa$.  For $\kappa \rightarrow \infty$, the entropy of the ground state 
tends to a value of unity, regardless of the value of $g_{1D}$ (provided $g_{1D}
> 0$).   In this limit the ground state of the system is described by one 
particle in each half of the double-well potential, and corresponds to the
Mott-insulator regime.  On the other hand, the initial value of $S$ (when
$\kappa = 0$) is sensitive to $g_{1D}$.  The higher the value of $g_{1D}$, the 
larger is the initial value of $S$.  As $S \rightarrow 1$ in the insulator
limit, it follows that the entropy of the ground state exhibits a less dramatic
variation with $\kappa$, for larger values of interaction strength.  For all of
the symmetric eigenstates, i.e. ground (solid line), second-excited (dot-dash 
line) and third-excited (dotted line), as the interaction strength is increased
the entropy of the eigenstates, in general, increases, consistent with Fig. 
\ref{fig:entropy_v_int_excited}.  In particular, the entropy of these symmetric 
states in the absence of a barrier ($\kappa = 0$), increases with increasing 
$g_{1D}$.  The second-excited state (dot-dash line) exhibits an entropy that 
monotonically increases with $\kappa$ for the range of parameter space 
considered.  By contrast, the third-excited state (dotted line) exhibits an 
entropy that both increases then decreases with raising of the barrier. 

One will also note that in the limit of large barrier heights (i.e. $\kappa
\rightarrow \infty$), the entropy of the second- and third-excited states tend
to the same value.  
Although not obvious from Fig. \ref{fig:entropy_v_barrier_excited}(d),
this fact has also been verified for the case of $g_{1D} = 10$.  Once again, a
handle on why this happens can be obtained from the wavefunction plots for these
eigenstates in Figs. \ref{fig:psi02} and \ref{fig:psi03}.  One can see that in
the presence of finite interactions, these two eigenstates become identical in
the insulator limit, except for some phase (compare row 4 of these figures). 
Fig. \ref{fig:entropy_v_barrier_excited} also suggests the the value of $S$ to 
which these two states converge, in the insulator limit, is greater than one and
increases with increasing interaction strength, $g_{1D}$.  

This behavior of the entropy may be qualitatively understood as follows.  Both
the second- and third-excited states correspond, in the insulator limit, to the
physical situation of two particles coexisting in either the right-well or the 
left-well.  As such, these states may be roughly represented by Bell-type states
of the form $\; 1/\sqrt{2} \left( \left. \left| 20 \right. \right> \pm \left.
\left| 02 \right. \right> \right)$ - see Sec.
\ref{subsubsect:bose_hubbard_entropy} for the definition of these basis states. 
Such a Bell state carries one e-bit of entanglement, with a corresponding von
Neumann entropy of unity.  However, beyond this, there are also correlations
between the two particles coexisting in the same well.  As can be seen, for
example, from Fig. \ref{fig:psi03} (4,4).  Here the repulsive interactions
between the particles occupying the same well leads to a partition of the
wavefunction, within each well, into two lobes.  Considering the double lobe seen
in the upper-right quadrant, corresponding to both particles co-existing in the
right well of the double-well.  The upper half of the lobe represents the
situation where particle 1 is on the left of this well and particle 2 is on the
right, the lower half-lobe corresponds to the reverse of this situation ($x_{1} 
\leftrightarrow x_{2}$).  In this case the correlations in the system are 
analogous to the correlations that are observed for the ground state, $\Psi_{0}$,
in the absence of any barrier ($\kappa = 0$).  These correlations (and therefore
$S$) are seen to increase as the interaction coupling is increased.  One 
significant distinction exists between these `single-well' correlations, seen in 
states $\Psi_{2,3}$, and the correlations seen in the ground state, for $\kappa =
0$.  On increasing $\kappa$, the second- and third-excited states tend to become
more confined and the two-particle wavefunction becomes increasingly localized in
the single-well.  However, the particle-particle interactions will compete with 
this effect, attempting to keep the two-particle wavefunction spread in space
and, in particular, minimized along the line $x_{1} = x_{2}$.  For the ground 
state this particular type of single-well competition between $\kappa$ and
$g_{1D}$ is not experienced.  So, in the insulator limit, the second- and 
third-excited states will have correlations arising from the realization of the
Bell-type state, and the `single-well' correlations due to the two interacting 
particles coexisting in the same well.  This combination of factors leads to an 
entropy which is greater than unity, with the contribution of the `single-well' 
correlations, in general, increasing with increasing interaction strength.

\subsection{\label{subsect:stimulating_excitations} Stimulating two-particle
excitations}
\begin{figure}[!ht]
      \begin{center}  
         \includegraphics[scale = 0.3]{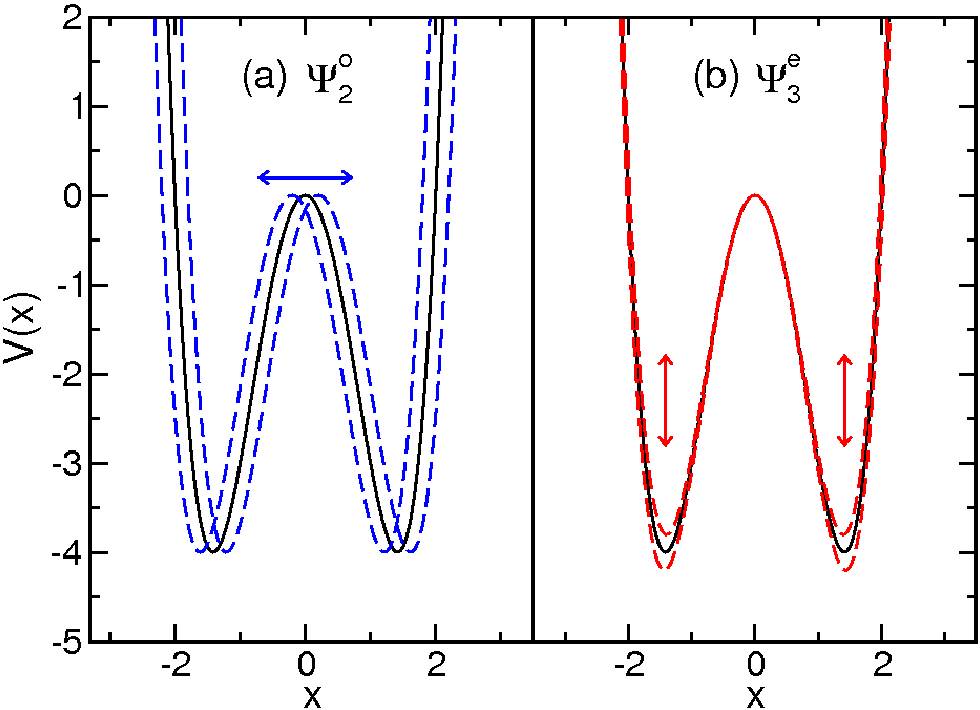} 
	 \includegraphics[scale = 0.26]{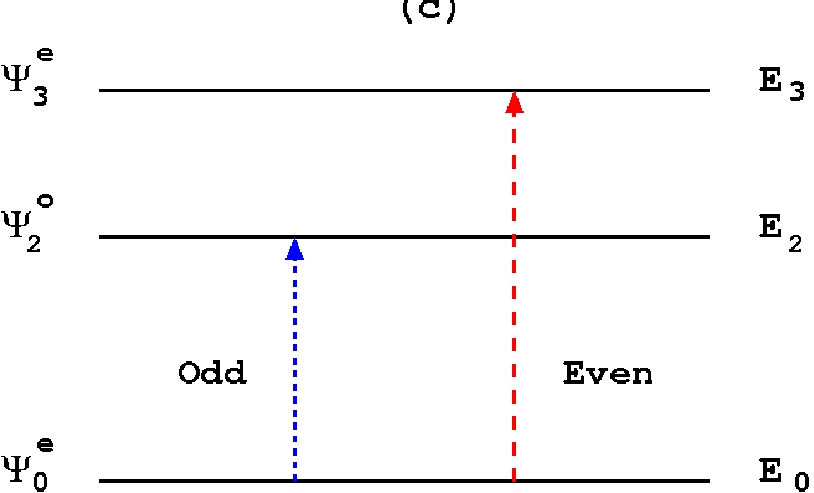}
      \caption{\label{fig:bhm_manipulations} Time-dependent manipulation of the
      two-particle state may be achieved by (a) oscillation of the trap from
      side-to-side or (b) the modulation of the barrier height.  The former 
      process provides a dipole coupling between the ground state, $\left. \left|
      \Psi_{0}^{\textrm{e}} \right. \right>$, and the state $\left. \left|
      \Psi_{2}^{\textrm{o}} \right. \right>$.  The second scheme will provide a 
      quadrupole coupling between the ground state and the state $\left. \left|
      \Psi_{3}^{\textrm{e}} \right. \right>$.  (c) shows a schematic energy level
      representation for these processes.}
      \end{center}
   \end{figure}

The previous results have clearly illustrated that manipulations of this 
two-particle system can be achieved through the variation of the control 
parameters $g_{1D}$ and $\kappa$ in some adiabatic manner.  However, one could 
also consider time-dependent manipulation of the state.  Considering the 
insulator limit, one may propose two methods of coupling these lowest levels: (a)
shaking the trap from side-to-side (b) modulating the barrier height (see Fig.
\ref{fig:bhm_manipulations}).  To first-order, the former represents a dipole 
excitation, capable of coupling $\left. \left| \Psi_{0}^{\textrm{e}} \right.
\right>$ and $\left. \left| \Psi_{2}^{\textrm{o}} \right. \right>$.  The latter 
scheme (to first-order) corresponds to a quadrupole excitation, capable of 
coupling states $\left. \left| \Psi_{0}^{\textrm{e}} \right. \right>$ and $\left.
\left| \Psi_{3}^{\textrm{e}} \right. \right>$.  In this way, by employing such 
techniques it should prove possible to exploit these three lowest eigenstates 
in order to engineer the two-particle state in a time-dependent fashion. 

Further investigation of this idea of time-dependent manipulation of the 
two-particle state could prove a useful extension to the present study.  In 
particular, a combination of time-dependent excitation processes and the 
adiabatic variation of control parameters, $g_{1D}$ and $\kappa$, should permit 
an impressive degree of control over the two-particle state, within this 
system.

\section{\label{sect:summary} Summary}
The system of two interacting particles in a prototypical double-well potential
of the form $V \left( x \right) = A \left[ x^{4} - \kappa x^{2} \right]$ has 
been considered.  Using a cartesian DVR, the eigenspectrum for this system has
been studied and the four lowest eigenstates have been obtained and 
investigated for varying barrier height and interaction strength.  For each 
state the two-particle eigenfunction, the momentum distribution and the von 
Neumann entropy have been examined.  It was found that the ground state
for this double-well system exhibits behavior that closely resembles that 
observed in a previous study of the $\delta$-split trap potential, \cite{murA07}.  
In particular, the ground-state wavefunction is suppressed along the lines $x_{1}
= 0$ and $x_{2} = 0$ as barrier height is increased, leading to a quadrant 
separation of the wavefunction.  In the presence of repulsive interactions 
($g_{1D} > 0$) only the contributions in the off-diagonal quadrants remain in 
the insulator limit ($\kappa \rightarrow \infty$).  In this limit the ground 
state of the system is composed of one particle in each half of the double-well.
The momentum distributions display an initial narrowing with increasing barrier 
height but with a broadening and high-energy wings being observed in the 
insulator limit.  Furthermore, the secondary peaks observed in the momentum 
distribution for the double well, in the non-interacting regime, are quickly 
suppressed in the presence of repulsive interactions.  The variation in the von 
Neumann entropy ($S$) with interaction strength shows remarkably similar 
behavior.  In all cases $S = 0$ in the absence of interactions and for $g_{1D} 
\rightarrow \infty$, $S$ saturates at a value close to unity.  Increasing the
height of the barrier, in each case, has the effect of making the entropy more 
sensitive to changes in the interaction strength, around $g_{1D} = 0$. 
Similarly, the behavior of the entropy with varying barrier height exhibits 
generic features between the two double-well systems.  In both cases the
ground-state entropy saturates at a value of unity as $\kappa \rightarrow 
\infty$, regardless of the value of $g_{1D}$.  The initial value of $S$ (i.e. 
the value of $S$ for $\kappa = 0$) is determined by the strength of the 
interaction, with larger interaction coupling leading to larger initial entropy.
As such, the sensitivity of $S$ to $\kappa$ is reduced for double-well systems 
with larger interaction couplings ($g_{1D}$).  This behavior of the ground-state
entropy is also illustrated within a Bose-Hubbard model, wherein the controllable
parameters are the on-site interaction ($U$) and the tunnelling strength ($J$).  

As well as examining the ground state of this double-well system, some of 
the properties of the three lowest excited states have also been studied.  Two 
of these states are found to be symmetric whilst one is antisymmetric and they
constitute the lowest band of the two-particle, double-well system.  The
antisymmetric state is found to be completely independent of the interaction
parameter ($g_{1D}$).  However, this state displays a dependence on the barrier 
height and in the limit of a high barrier becomes degenerate with the ground 
state - corresponding, physically, to the situation of each particle residing in 
a separate, isolated well.  The von Neumann entropy for this antisymmetric state 
is identically equal to one for all $\kappa$ and $g_{1D}$.

The second- and third-excited states are symmetric.  In the insulator limit
(provided $g_{1D} > 0$) the states become degenerate and correspond to the
physical situation where both particles occupy the same well.  Both eigenstates
demonstrate momentum distributions that are double-humped, with the double-hump 
giving way to a single peak in the insulator limit.  For $g_{1D} = 0$ the 
third-excited state exhibits secondary peaks in the momentum distribution, 
similar to the ground state in the non-interacting regime.  The entropy of both 
states increases with $g_{1D}$, with that of the third-excited state showing a 
more marked variation.  As for the ground state, increasing barrier height 
$\kappa$ has the effect of increasing the sensitivity of the entropy to 
variations in $g_{1D}$ (about $g_{1D} = 0$).  In the insulator limit the entropy
of the third excited state is found to follow, almost identically, that of the 
ground state for small $g_{1D}$.  As the ground state entropy saturates at $S 
\approx 1$, the entropy of the third excited state continues to increase and, 
for larger $g_{1D}$, follows, almost identically, that of the second-excited 
state.  Indeed, in the insulator limit and for fixed interaction strength, the 
second- and third-excited states are found to have the same entropy (as follows 
from the physical equivalence of these states in this limit).  The entropy, in 
this case, is proposed to have two contributions due to (i) the realization of 
a Bell-type state with both particles co-occupying either the left \textit{or} 
right well (ii) single-well correlations, owing to the repulsive interaction of 
the two particles occupying the same well.

\subsection{\label{subsect:outlook} Outlook}
The double-well arrangement studied in this work represents a more
experimentally realizable system, compared to the $\delta$-split trap previously
considered.  Having characterized the properties of the ground and
lowest-excited eigenstates, the foundation is laid for future investigation 
into state manipulations using this system.  Future avenues may include the
time-dependent manipulation of states through shaking of the trap, an 
oscillating barrier height or introduction of a constant, or oscillating, field 
gradient.  These time-dependent manipulations, along with the adiabatic variation
of the control parameters $g_{1D}$ and $\kappa$, should allow for comprehensive
state engineering within the lowest band of this two-particle system.

\begin{acknowledgements}
The authors would like to thank John Goold, Thomas Busch and Mauro Paternostro
for helpful discussions.  DSM would like to acknowledge funding from the 
Department for Employment and Learning (NI) and the support of the Sorella Trust
(NI).
\end{acknowledgements}

\bibliography{two_part_double_well}

\end{document}